\documentclass[twocolumn,times]{aastex631}

\usepackage{amsmath}
\usepackage{mathtools}
\usepackage{eucal}
\usepackage{amssymb}
\usepackage{enumitem}
\usepackage{svg}
\usepackage{comment}

\usepackage{placeins}

\definecolor{webgreen}{rgb}{0,.5,0}
\definecolor{webbrown}{rgb}{.6,0,0}

\usepackage[toc,page]{appendix}

\def \beq{\begin{equation}}
\def \eeq{\end{equation}}
\def \bea{\begin{eqnarray}}
\def \eea{\end{eqnarray}}

\received{\today}

\newcommand{\TKep}{\mbox{$ T_{\rm Kep} $}}
 
\newcommand{\Tbb}{T_{\rm 2b}}

\newcommand{\vesc}{v_{\rm esc}} 

\newcommand{\Mbh}{M_{\bullet}}
\newcommand{\lnL}{\ln{\Lambda}}
\newcommand{\lnLo}{\lnL_0}

\newcommand{\p}{\partial}
\newcommand{\rmd}{\mbox{${\rm d}$}}

\newcommand{\Msun}{\mbox{$ M_\odot $  }}

\newcommand{\rh}{\mbox{$r_{\rm h}$}}

\newcommand{\mstar}{m_{\star}}

\newcommand{\Nf}{N_{\rm f}}

\newcommand{\agw}{a_{\rm gw}}

\newcommand{\ac}{a_{\rm c}}
\newcommand{\aco}{a_{\rm c 0}}
\newcommand{\rpgw}{r_{p,{\rm gw}}  }

\newcommand{\Tgw}{T_{\rm gw}}

\newcommand{\bninty}{b_{\rm 90}}
\newcommand{\bej}{b_{\rm ej}} 
\newcommand{\vej}{v_{\rm ej}}

\newcommand{\Pdot}{\dot{P}_{\rm ej}}
\newcommand{\Pdotav}{\langle {\dot{P}}_{\rm ej} \rangle}
\newcommand{\Nd}{N_{\rm d}}
\newcommand{\gamd}{\gamma_{\rm d}}
\newcommand{\Nrefd}{N_{\rm f d}}

\newcommand{\fbh}{f_{\rm bh}} 
\newcommand{\aref}{a_{\rm f}}
\newcommand{\Nref}{N_{\rm f}}
\newcommand{\Fej}{F_{\rm ej}}
\newcommand{\scN}{\mathcal{N}}
\newcommand{\scR}{\mathcal{R}}

\newcommand{\Mmw}{M_{\bullet {\rm mw}}}
\newcommand{\Int}{\mathcal{I}}
\newcommand{\Feju}{\Fej^{({\rm u})}}
\newcommand{\Fejd}{\Fej^{({\rm d})}}
\newcommand{\fgw}{f_{\rm gw}}
\newcommand{\adot}{|\dot{a}|}
\newcommand{\Rdot}{|\dot{\scR}|}
\newcommand{\Nej}{\dot{N}_{\rm ej}}

\begin{document}

\title{Extreme-mass ratio inspirals in strong segregation regime - to inspiral or to get ejected?}

\correspondingauthor{Karamveer Kaur}
\email{karamveerkaur30@gmail.com}

\author[0000-0001-6474-4402]{Karamveer Kaur}
\affiliation{Technion - Israel Institute of Technology, Haifa, 3200002, Israel}

\author[0000-0002-5004-199X]{Hagai Perets}
\affiliation{Technion - Israel Institute of Technology, Haifa, 3200002, Israel}

\shorttitle{EMRI rates for strongly-segregated cusps}
\shortauthors{Kaur and Perets}

\begin{abstract}

Extreme-mass ratio inspirals (EMRIs) of stellar-mass black holes (BHs) are among the main targets for upcoming low-frequency gravitational wave (GW) detectors such as the Laser Interferometer Space Antenna (LISA).
 In the classical scenario, EMRIs are formed when BHs scatter off each other and are driven onto highly eccentric orbits that gradually inspiral due to GW emission. If the cluster is in a state of strong mass segregation, the BHs are expected to reside in a steep cusp around the central massive black hole (MBH), which would facilitate more efficient EMRI formation. However, strong mass segregation may also lead to an increased rate of ejections due to close encounters between the BHs. Here, we test the relevance of such ejections for EMRI formation by numerically solving a two-dimensional Fokker-Planck equation. Our formalism includes the effects of two-body relaxation, GW dissipation, and ejections. We find that the EMRI formation rate can be suppressed due to ejections by more than an order of magnitude for strongly segregated BH cusps with density index $\gamma\gtrsim 2.25$ around central MBHs of mass $\Mbh \lesssim 10^6 \Msun$. The EMRI formation rate levels off up to a maximum value of $\simeq 200~{\rm Gyr}^{-1}$ due to ejections, which is roughly an order of magnitude lower than the usual scenarios ignoring ejections for steep BH cusps around low mass MBHs. Our analysis brings forth the significance of strong scatterings for EMRI formation in galactic nuclei.  

\end{abstract}

\keywords{Galaxy nuclei (609), Galactic center (565), Stellar dynamics (1596), Supermassive black holes (1663), Stellar mass black holes (1611), Gravitational wave sources (677)}

\section{Introduction}

In the near future, space-based gravitational wave observatories like LISA and TianQin will usher in a new era of mHz gravitational wave astronomy, providing a window into the galactic nuclei harboring massive black holes (MBHs) \citep{eLISA_Consort2013,Mei_2021,Kormendy_Ho_2013}. Many of these MBHs reside within dense nuclear star clusters (NSCs), dynamic environments where a multitude of astrophysical phenomena can occur, including the formation of extreme-mass ratio inspirals (EMRIs) \citep{Alexander_2017}. EMRIs, consisting of stellar-mass black holes (BHs) captured in close orbits around MBHs, are key targets for this low-frequency gravitational wave (GW) detectors \citep{Amaro-Seoane_2007}.

The classical EMRI formation channel posits that two-body (2B) scatterings drive BHs into highly eccentric orbits, which subsequently undergo inspiral due to GW emission \citep{Hils_1995,Sigurdsson_1997,Freitag_2001,Hopman_Alexander_2005}\footnote{There is a variety of non-2B EMRI formation channels which may operate under specific conditions \citep{Miller_2005,Perets2007,Levin_2007,Bode_2014,Pan2021,Raveh_2021,Naoz_2022}, but here we focus on the classical scenario.}. However, theoretical predictions of EMRI rates remain uncertain, varying by up to two orders of magnitude \citep{Hopman_Alexander_2005,Amaro_Seoane_Preto_2011,Merritt_2015,Bar-Or_2016,Aharon_Perets_2016,Raveh_2021,KaurFP_24,Rom_24}. This uncertainty stems from our incomplete knowledge of the internal structure of NSCs, in particular, the distribution and number density of stellar BHs, both of which significantly influence relaxation timescales and the rate of BH influx into the GW loss cone \citep{Amaro-Seoane_2018}.

Fokker-Planck investigations dealing with scattering-driven energy relaxation have sought to elucidate the structure of NSCs theoretically. \citet{Bahcall_Wolf_1977} identified a zero-flux solution characterized by a density profile index $\gamma$=7/4 for the dominant stellar population. However, \citet{Alexander_Hopman2009} proposed an alternative solution branch with non-zero net flux, leading to steeper density profiles with $\gamma\simeq2-11/4$, known as the strong segregation regime. This regime is particularly relevant when BHs constitute a minority population compared to stars. In this scenario, the more numerous stars establish a shallower profile consistent with the zero-flux solution, while the BHs, experiencing dynamical friction against the stellar background, form a steeper cusp (\citealt{Alexander_Hopman2009}; see \citealt{Linial_Sari_22} for a different perspective). The appeal of the strong segregation regime lies in its ability to produce high EMRI rates, on the order of hundreds per Gyr for a Milky Way-like galaxy, even with physically plausible BH number fractions as low as 10$^{-3}$ \citep{Amaro_Seoane_Preto_2011}. These high rates are facilitated by the steep BH density profiles, which yield high BH densities in the inner regions ($\approx0.01 \rh$; $\rh$ being the radius of influence of central MBH) where EMRIs predominantly form \citep{Hopman_Alexander_2005}, and also shorten the relaxation timescales associated with EMRI formation.

However, the increased BH densities in strongly segregated cusps raise the possibility of strong 
encounters that could eject stars and compact objects. Such ejection effects have been previously 
explored in the context of globular clusters \citep{Henon_1960,Lin_Tremaine_1980,Goodman_1983}, 
where they are important for cluster evaporation. In NSCs, such strong encounters can eject stars before they achieve sufficiently eccentric orbits to interact with the MBH strongly. In particular, \cite{Teboul_2024} showed that the ejections of stars on high eccentricity ($e$) orbits can significantly quench the rates of tidal disruption events (TDEs) in strongly segregated NSCs. This brings out the significance of strong scatterings for high $e$ orbits, particularly relevant for general loss cone dynamics around an MBH \citep{Weissbein_Sari_2017,Teboul_2024}. Indeed, strong scatterings can induce substantial changes in BH orbital parameters \citep{Agekyan_1959,Henon_1960,Henon_1960b,Ashurov2004,Henon11,Teboul_2024}, potentially ejecting BHs from the highly eccentric orbits necessary for EMRI formation, especially in the strongly segregated NSCs which have enhanced BH densities in their inner regions.

In this study, we examine the impact of strong scattering-driven ejections on EMRI formation within strongly segregated BH cusps. We evaluate a numerical solution of a two-dimensional Fokker-Planck equation that simultaneously accounts for 2B relaxation, GW-induced orbital decay, and ejections triggered by strong scatterings. For the first time in consideration of strong scatterings, we take into account the depletion of the inner cusp of (BH) scatterers due to GW loss cone. 
This prevents a significant overestimation of ejection rates due to artificially high densities in the inner regions of the cluster. We employ the Fokker-Planck solution to calculate time-dependent EMRI rates and assess the level of suppression induced by ejections. It turns out that strong scatterings can lower the EMRI rates up to 1-2 orders of magnitude in strongly segregated cusps with steepest profiles ($\gamma \gtrsim 2.25$) around MBHs with low mass $\Mbh \approx 10^{5-6} \Msun$ (table~\ref{tbl_rates}, figure~\ref{fig_rates_diff_para}). This is especially interesting given the maximum LISA sensitivity for EMRI detections around these low-mass MBHs \citep{Babak2017,Rom_24}. 

This paper is structured as follows. In Section~\ref{sec_phy_setup}, we delineate the relevant physical processes and present our Fokker-Planck formalism. Section~\ref{sec_results} discusses the numerical solution of the equation and presents the resulting EMRI rates. Finally, in Section~\ref{sec_discus} we summarize our findings and discuss their astrophysical implications.

\section{Physical set-up} 
\label{sec_phy_setup}

We consider a single BH population of mass $m$ forming a compact power-law density profile, with index $\gamma$, well inside the radius $\rh$ of influence of the central MBH of mass $\Mbh$. The number profile $N(a) = \Nref (a/\aref)^{3-\gamma}$ represents the BHs with semi-major axes $\leq a$, such that $\Nref = \fbh (\Mbh/\mstar)$ is the total number of BHs within the reference radius $\aref \simeq 0.1 \rh$ \citep{Amaro_Seoane_Preto_2011}. Here, stars of mass $\mstar$ form the most abundant stellar population within $\rh$, and $\fbh$ is the number fraction for BHs relative to stars. In spite of the presence of more abundant lighter components (for e.g. stars), the high number density of BHs in these inner regions makes them the dominant scatterers, for both weak scatterings responsible for angular momentum ($L$) relaxation, and strong scatterings leading to ejections. 

In the classical picture of EMRI formation, many random weak or 2B scatterings may ultimately lead a BH onto a highly eccentric orbit, such that it loses its orbital energy and gradually shrinks in semi-major axis ($a$) due to GW emission upon pericentric interactions with MBH. However, as BHs attain these high eccentricity $e$ orbits, they become susceptible to ejections owing to a possible close encounter with another BH near its pericenter. This can lead to suppression in the formation rates of EMRIs. We discuss these physical phenomena below and implement them in a Fokker-Planck (FP) framework to study the evolution of BH distribution.

\subsection{Two-body scatterings} 

Due to numerous 2B scatterings among each other, BHs undergo a random walk in two-dimensional space of energy $\varepsilon = G \Mbh/(2 a)$ and angular momentum $L = \sqrt{2 G \Mbh r_p}$ over 2B relaxation time $\Tbb$; $r_p$ being the pericentric distance from central MBH. While $\Tbb$ represents a reference time for near-circular orbits, relaxation in $r_p$ for highly eccentric orbits, relevant for EMRI formation, occurs over much shorter timescale $\Tbb^{L} = (2 r_p/r) \Tbb$ compared to relaxation in $a$ \citep{Binney_Tremaine}. Hence, we neglect the relaxation in $a$ and consider a fixed spatial profile $N(a)$ of scatterers at all times. We employ the following form of 2B timescale \citep{Merritt_2011,Merritt_2013,Bortolas2019}:   
\beq 
\Tbb(a) = \frac{3\sqrt{2} \pi^2}{32 C} \frac{\TKep(a)}{\lnL} \bigg( \frac{ \Mbh}{m} \bigg)^2 \frac{1}{N(a)} \,
\label{Tbb}
\eeq 
with dynamical timescale $\TKep = \sqrt{a^3/(G \Mbh)}$, and Coulomb logarithm $\lnL \simeq \log{(\Mbh/m)}$. The factor $C$ has a weak dependence on $\gamma$ (see for e.g. appendix A of \citealt{Bortolas2019}), and we use $C = 1.35$ corresponding to a $\gamma = 7/4$ Bahcall-Wolf (BW) cusp \citep{KaurFP_24}. 
Similar to previous studies \citep{Merritt_2013}, the inclusion of 2B scatterings in FP equation~\ref{FP} considers that: (1) background scatterers (BHs themselves in our scenario) follow a thermal distribution, and (2) drift and diffusion coefficients arising from $L$-relaxation are combined using fluctuation-dissipation theorem to get an effective diffusion term in $r_p$ \citep{Lightman_Shapiro1977}. Further, we neglect resonant relaxation \citep{Rauch_Tremaine_1996}, as it has been found to be ineffective in influencing overall EMRI rates \citep{Alexander_2017RR}.    

\subsection{GW-induced dissipation}

As a fraction of BHs attain high $e$, they can come sufficiently close to MBH during pericentric passages leading to energy and angular momentum losses by GW emission. The resulting orbital shrinkage or inspiral occurs over a GW timescale $\Tgw$ given explicitly as \citep{Peters_1964,KaurFP_24}:
\beq 
\Tgw(r,r_p) \equiv \frac{a}{|\dot a|} = \frac{96 \sqrt{2} }{85} \frac{R_s}{c} \frac{\Mbh}{m} \bigg( \frac{r_p}{R_s} \bigg)^4 \sqrt{\frac{a}{r_p}}\,,
\label{Tgw}
\eeq 
as suited for high $e$ orbits. Here $R_s= 2 G \Mbh/c^2$ is the Schwarzschild radius. We take into account this orbital shrinkage due to energy and angular momentum losses in the FP equation~\ref{FP}.

For high $e$ orbits with $\Tgw \leq \Tbb^{L}$, orbital inspiral dominates over the stochastic effect of 2B scatterings, and orbit eventually becomes an EMRI. This condition effectively presents a GW loss cone, for $r_p \leq \rpgw$, with the critical periapsis $\rpgw$ given explicitly as:
\beq 
\rpgw(a) = R_s \bigg[   \frac{\xi}{\lnL}  \frac{\mstar}{m}  \frac{1}{\fbh} \bigg]^{2/5} 
           \bigg( \frac{\aref}{a}\bigg)^{2(3 - \gamma)/5}  
\label{rpgw}
\eeq 
where $\xi \simeq 1.7$ \citep{KaurFP_24}. 

The critical semi-major axis $\ac$ roughly demarcates the outer boundary within which most EMRIs occur \citep{Hopman_Alexander_2005}, because GW loss cone for $\ac$ coincides with the capture radius $\simeq 4 R_s$ \footnote{The capture radius of $4 R_s$ corresponds to the most bound orbit around a Schwarzschild MBH, with an angular momentum $L = 4 G \Mbh/c$.}. Using $\rpgw(\ac) = 4 R_s$, we get:
\beq 
\ac = \aref \bigg[   \frac{\xi}{32 \lnL}  \frac{\mstar}{m}  \frac{1}{\fbh} \bigg]^{1/(3-\gamma)} \,.
\label{ac}
\eeq 
This implies $\ac/\rh \simeq 0.03-0.06$ for $\gamma = 7/4 - 5/2$ using $\aref = 0.1 \rh$, $\lnL = 10$, $m = 10 \mstar$, $\fbh = 10^{-3}$.  Outside $\ac$, most highly eccentric orbits lead to direct plunges into the MBH. However, for a central intermediate-mass black hole ($\Mbh \lesssim 10^4 \Msun$), this boundary becomes somewhat blurred and a fraction of EMRIs may arise from $a \gg \ac$ \citep{Qunbar_Stone_2023}.   

Further, for highly bound orbits $a \leq \agw$, GW-induced orbital inspiral begins to dominate even for circular orbits. Using $\agw = \rpgw(\agw)$, we have:
\beq 
\agw = R_s \bigg[   \frac{\xi}{\lnL}  \frac{\mstar}{m}  \frac{1}{\fbh} \bigg( \frac{\aref}{R_s} \bigg)^{(3-\gamma)} \bigg]^{1/(11/2-\gamma)} 
\label{agw}
\eeq 
which gives $\agw \simeq 23-170~R_s$ for $\gamma = 7/4 - 5/2$ \footnote{Both $\agw$ and $\ac$ are decreasing functions of $\gamma$, as higher BH density resulting from higher $\gamma$ implies the stronger influence of scatterings and whole GW regime shifts closer to MBH.}, and $\Mbh = 4 \times 10^6 \Msun$ and $\rh = 2$pc corresponding to our Galactic center \citep{Ghez_2008,Gillessen_2009}.   
This roughly marks the inner edge till where BH cusp defined by power-law density index $\gamma$ may sustain, while inside this region the cusp is depleted due to GW losses.

\subsection{Strong scatterings}

For highly eccentric orbits, ejection from the system is the most probable outcome of strong scatterings. In fact, it might not even require \emph{strong} scatterings (with impact parameter $\bninty = G m/v^2$ such that $\Delta v \sim v$), since the test BH is moving near escape speed $\vesc \simeq \sqrt{2 G \Mbh/r}$ for most phases of its orbit satisfying $r \ll a$. Only a small velocity impulse is needed for its ejection $\Delta \vej = \vesc - v = \vesc \, r/(4 a) \ll v \sim \vesc$. This corresponds to a large ejection impact parameter $\bej = G m/(v \Delta \vej) \simeq 2 a \, m/\Mbh \gg \bninty$, making ejections likely for high $e$ orbits as they navigate through high-density regions of inner BH cusp during the pericentric passage.   

\cite{Henon_1960,Hen+69} calculated the local probabilistic rate of ejection $\Pdot$ of a test star for general stellar clusters, and more recently \cite{Teboul_2024} evaluated it for a Keplerian potential. This gives the ejection probability $\Pdot \approx  \pi \bej^2 v {N(r)}/{(4 \pi r^3)}$ of a test star with Keplerian orbital parameters $\{a,r_p \}$ near phase $r$ of its orbit; equation~\ref{Pdot_fin} gives an exact form of $\Pdot$. The actual time spent by a star at the phase $r$ is proportional to $(r/a)^{3/2}$. This implies an orbit-averaged ejection rate of $\Pdotav \approx (\TKep)^{-1} \int \rmd r \, \Pdot/v \approx \Pdot (r/a)^{3/2}$, which gives:
\beq 
\Pdotav \! \simeq \! \frac{3\sqrt{2} \pi}{32 C} \frac{1}{ \lnL \, \Tbb(\aref)} \frac{\Gamma(\gamma+1)}{\Gamma{(\gamma+1/2)}}  \sqrt{\frac{a}{\aref} } \frac{r_0}{\aref} \bigg( \frac{\aref}{r_0}  - \frac{\aref}{2 a} \bigg)^{\gamma}  
\label{Pdotav_approx1}
\eeq 
see appendix~\ref{app_ej_rate} for the details of derivation. The dominant contribution to ejections comes from the phase $r_0 = {\rm max}[r_p,\agw]$. This is because ejections due to strong scatterings dominate for small $r$ phases of an orbit near its pericenter, which is in contrast to the case of $L$-relaxation owing to weak scatterings \citep{Binney_Tremaine}.

\emph{Depletion of inner cusp}: Here we account for the depletion of inner BH cusp by GW loss cone, because dense parts of BH cusp cease to exist within $\approx \agw$, marking the inner edge of the depleted cusp. This depletion radius corresponds to a few 100 $R_s$ for most realistic  galactic nuclei with $\Mbh \simeq10^{4-7}\Msun$. Hence, it is important to account for this depletion by choosing the appropriate phase $r_0$ that contributes maximally to ejections, as we consider above. This effect of depletion of inner cusp by GW loss cone has not been accounted in previous studies, and might have led to overestimation of ejection rates.

Further, rather than using the above approximate expression for $\Pdotav$, we evaluate numerically the integral in equation~\ref{Pdotav_gen} for an exact expression of $\Pdotav$, which is then employed in the Fokker-Planck framework described below. The exact evaluation of $\Pdotav$, as detailed in the appendix~\ref{app_ej_rate}, takes into account the chosen (physically motivated) form of orbital configuration (equation~\ref{Nd_depleted}) of depleted region $a \lesssim \agw$. But, it is in close agreement with the above approximate form of $\Pdotav$ that just depends on the radius of depletion $\agw$ \footnote{Also, both cases lead to the Fokker-Planck solution and resulting EMRI rates which are compatible upto a few tens of percent. This shows that results are not sensitive to the exact form of depleted orbital configuration inside $\agw$, that does not contribute much to ejections. }.   

\subsection{Fokker-Planck framework}

 We track the evolution of 2D density $\scN$ of BH distribution in $\{a , \scR \}$-plane with $\scR = 1- e^2 = 1 - (1-r_p/a)^2$, because these coordinates  provide a suitable rectangular grid for the numerical solution \citep{Merritt_2013}. We include the influence of all three processes described above into the following Fokker-Planck (FP) equation:  
 \beq 
 \begin{split}
& \frac{\p \scN}{\p t} = \frac{1}{ a^{\gamma-3/2}} \frac{\p}{\p \scR}\bigg( \scR \frac{\p \scN}{\p \scR} \bigg)  
\,+\, \frac{\p}{\p a}\bigg( \adot  \scN \bigg)\, \\[1ex]
& \qquad \quad -\, \frac{\p}{\p \scR}\bigg( \Rdot  \scN \bigg)
\,-\, \Fej(a,\scR) \scN \,.
\end{split}
\label{FP}
\eeq
The first diffusion term on the right side of the above partial differential equation (PDE), refers to the relaxation in $\scR$ due to weak two-body scatterings \citep{Merritt_2013}. The second and third advection terms account for losses due to GWs, with GW advection speeds $\adot$ and $\Rdot$ respectively in $a$ and $\scR$ directions \citep{Peters_1964} given as: 
\beq 
\begin{split}
& \adot = \frac{96 \times 2^{7/2}}{425 \, a^3 \, \scR^{7/2} \, A} \, \bigg( 1 + \frac{73}{24} (1-\scR) +\frac{37}{96} (1-\scR)^2 \bigg)  \\[1ex] 
& \Rdot = \frac{304 \times 2^{7/2} (1-\scR)}{425 \, a^4 \, \scR^{5/2} \,  A} \, \bigg( 1 + \frac{121}{304} (1-\scR) \bigg) \,.
\end{split}
\label{GW_speeds}
\eeq 
Here the constant factor $A = \Tgw(\aref)/\Tbb(\aref)$ is the ratio of GW to 2B timescales (equations~\ref{Tbb} and \ref{Tgw}) for circular orbits of radius $\aref$. The fourth sink term depicts ejections due to strong scatterings, with the ejection factor \footnote{ Our approach is similar to \citet{Teboul_2024} who investigate the effect of strong scatterings on TDE rates, and include a similar sink term, without considering the depletion of inner BH cusp, in a Fokker-Planck formulation (excluding GW losses).  }:
\beq 
\begin{split}
 \Fej & \equiv \Pdotav \Tbb(\aref) \\[1ex]
  & \simeq \frac{3\sqrt{2} \pi}{32 C \lnL}  \frac{\Gamma(\gamma+1)}{\Gamma{(\gamma+1/2)}} \sqrt{a} \, r_0 \, \bigg( \frac{1}{r_0}  - \frac{1}{2 a} \bigg)^{\gamma} 
 \end{split}
 \label{Fej_ana} 
 \eeq 
  The above approximate form of $\Fej$ employs the simplified analytical expression for $\Pdotav$ in equation~\ref{Pdotav_approx1}. However, for implementation in our code, we generate tables for $\Fej$ (using equations~\ref{FejA} and \ref{FejB}) by numerically solving the integrals in the exact expression of $\Pdotav$ in equation~\ref{Pdotav_gen}. Here, we have described all lengths in the unit of $\aref$, and time $t$ in a unit of $\Tbb(\aref)$.  
  
  We solve the above PDE in the domain $a \in [\agw/\aref, 2 \ac/\aref]$ for an isotropic initial distribution (with $\scN$ independent of $\scR$), respecting the initial power-law profile $N(a) \propto a^{3-\gamma}$. This initial condition follows an empty (capture) loss-cone \footnote{The condition of empty loss cone for capture can become inaccurate for low-mass MBHs, where Browninan motion of central MBH can also become important \citep{Lin_Tremaine_1980,Merritt05,Merritt07}. However, we do not consider these effects in the current study for simplicity.}, such that $\scN = 0$ for $\scR \leq \scR_{\rm cap}(a) = {8 R_s}/{(a \, \aref)}$. Here, $\scR_{\rm cap}(a)$ defines a capture boundary corresponding to pericenter $r_p = 4 R_s$. The evolution of BH distribution $\scN(a,\scR)$ is followed till time $\Tbb(\ac)$, which is 2B relaxation time at the critical semi-major axis $\ac$ (equation~\ref{ac})  and is the timescale over which BHs channelling most of the EMRIs are expected to get depleted. 
  
  The upper boundary for circular orbits $\scR = 1$ has a zero in-flow of BHs, with $\p \scN/\p \scR |_{\scR = 1} = 0$. For outer-most grid column $a = 2 \ac$, we consider zero GW fluxes because this boundary lies well outside GW loss cone defined by equation~\ref{rpgw}. For grid cells just above the capture boundary $\scR_{\rm cap}(a)$, the GW flux in $\scR$ is not considered to avoid unphysical flow out of the empty loss cone, and only advective flux in $a$ maintains the GW-induced orbital inspiral.

   We consider a $50 \times 50$ rectangular and logarithmically uniform grid in $a \in [\agw/\aref , 2 \ac/\aref]$ and $\scR \in [\scR_{\rm cap}(2 \ac)/2,1]$ \footnote{We also check the convergence of our results by increasing the grid resolution to $100 \times 100$ for a subset of parameters, and find a good agreement for resulting suppression of EMRI rates upto a few per cent. }. 
   Our numerical scheme is based on mid-point discretization for diffusion term in $\scR$, and upwind scheme for GW advection terms so that physical direction of flow is maintained. We employ a forward Euler scheme for the time evolution, that respects Courant condition for numerical stability \citep{num_recipes_1992}.  We discuss the numerical solution and implications for resulting EMRI rates in the coming section.

\section{Results} 

\label{sec_results}

\begin{figure*}
    \centering
    \includegraphics[width=0.425\textwidth]{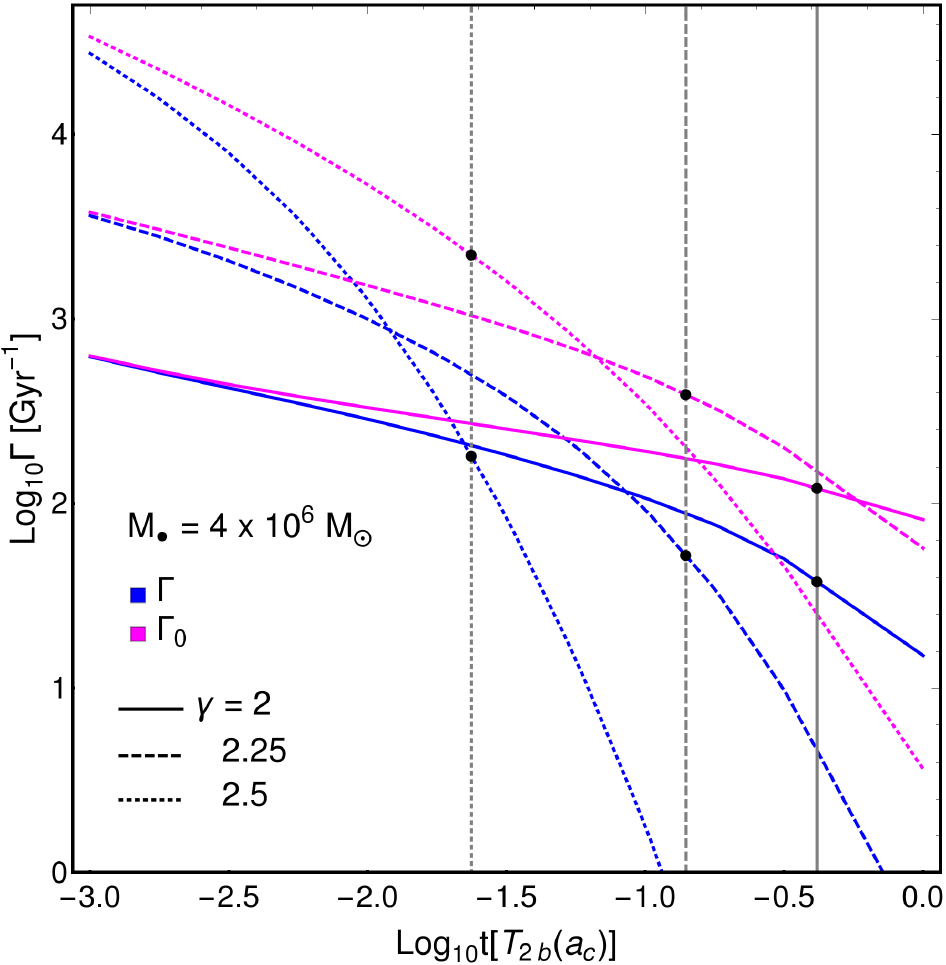}
    \hspace{1cm}
    \includegraphics[width=0.425\textwidth]{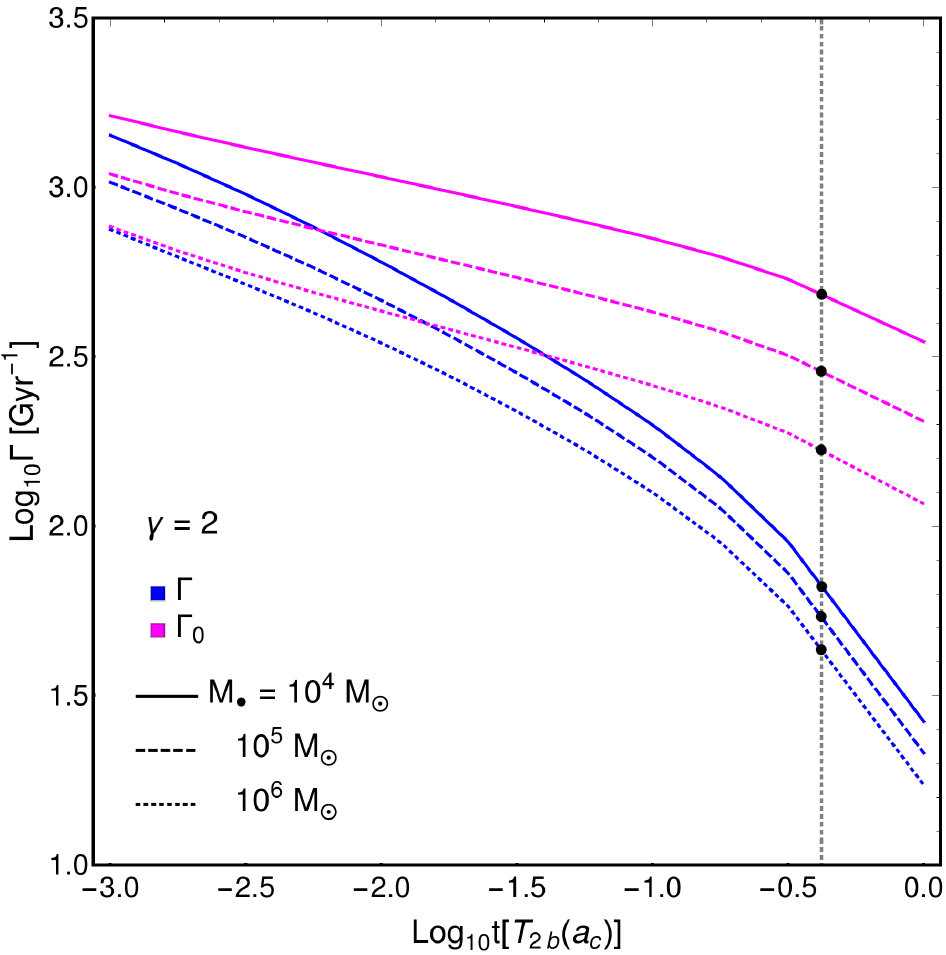}
    \caption{Evolution of EMRI rates, $\Gamma$ (including ejections, in blue) and $\Gamma_0$ (without ejections, in magenta), shown as function of time in units of $\Tbb(\ac)$ for various density slopes $\gamma$ of BH cusp. [Left panel] Rates for various BH density index $\gamma = 2 - 2.5$ are presented while fixing MBH mass $\Mbh = 4 \times 10^6 \Msun$. [Right panel] Rates for $\gamma = 2$ while varying $\Mbh = 10^{4-6} \Msun$. Evidently, the suppression is strongest for steepest cluster densities with higher $\gamma$ and/or lower MBH mass $\Mbh$. Further, the level of rate suppression due to ejections driven by strong scatterings, is time-dependent. The comparison at time $T_0 = \Tbb(\aco)$ (vertical gray lines) is apparently the most relevant in astrophysical settings that may sustain a steady state (see the text for more details). }
    \label{fig_rate_evo}
\end{figure*}

\begin{table*}
\centering
\footnotesize
{\renewcommand{\arraystretch}{1.2}
\begin{tabular}{|c| c|c| c|c|c|c|c|}
  \hline
$\gamma$  & $\Mbh/\Msun$ & $\ac/{\rm mpc}$ & $\aco/{\rm mpc}$ & $T_0/{\rm yr}$ & $\Gamma_0/{\rm Gyr^{-1}}$ & $\Gamma/{\rm Gyr^{-1}}$ & $\Gamma/\Gamma_0$    \\
  \hline 

1.75 & $4 \times 10^6$ &120  &   36   & $1.7 \times 10^9$  &  56     &  32   &  0.57 \\
   2  & $4 \times 10^6$ &110 &  18   & $7.8 \times 10^8$   &  120    &  38  &  0.31 \\ 
   2.25 & $4 \times 10^6$ & 87 &  6.3    & $1.9 \times 10^8$  &  390   &  52  &  0.13  \\  
    2.5 & $4 \times 10^6$ & 58 &  1.4   & $1.8 \times 10^7$ &  2200   &  180   & 0.08  \\  
    \hline 

 1.75 & $10^5$ &  19 & 5.8 & $1.7 \times 10^7$  &  130  &  62 & 0.49 \\
 2 &  $10^5$ &  17 & 3.0 &  $7.9 \times 10^6$  & 290  & 54 & 0.19 \\
 2.25 &  $10^5$ & 14 & 1.0 & $2.0 \times 10^6$  & 1100  & 40  & 0.037 \\
 2.5 &  $10^5$ & 9.1 &  0.20 & $1.6 \times 10^5$ & 8000 & 55 & 0.0069 \\
\hline


1.75 & $10^6$ & 61 & 18 & $3.0 \times 10^8$ &  75 &  40 &  0.54 \\
2 &   $10^6$ & 54 & 9.4 & $1.4 \times 10^8$ &  170 & 43 & 0.26 \\
2.25 & $10^6$ & 44 & 3.2 & $3.4 \times 10^7$ & 590 & 50 & 0.085 \\
2.5 &  $10^6$ &  29 & 0.7 & $3.2 \times 10^6$ & 3300 &  100 &  0.031 \\ 

\hline 


 1.75 & $10^7$ & 190 & 58 &  $5.3 \times 10^9$ & 45 & 27 & 0.59 \\
 2 & $10^7$ &  170 &  30 &  $2.5 \times 10^9$ & 97 & 34 & 0.35 \\
 2.25 & $10^7$ & 140 & 10 &  $6.2 \times 10^8$ & 300 & 53 & 0.18 \\
 2.5 & $10^7$ & 91 & 2.3 & $0.58 \times 10^8$ & 1500 & 200 & 0.13 \\

 \hline 

\end{tabular}
}
\caption{EMRI rates, $\Gamma_0$ (usual scenario without ejections) and $\Gamma$ (with ejections) from the numerical solution are quoted for various cluster density slopes $\gamma$ and MBH mass $\Mbh$. These are the instantaneous rates at time $T_0 = \Tbb(\aco)$. Here  $\aco$ is the numerically deduced maximum semi-major axis within which most EMRIs occur; while $\ac$ is its analytical counterpart as detailed in the text.   }
\label{tbl_rates}
\end{table*}

\begin{figure}
    \centering
    \includegraphics[width=0.45\textwidth]{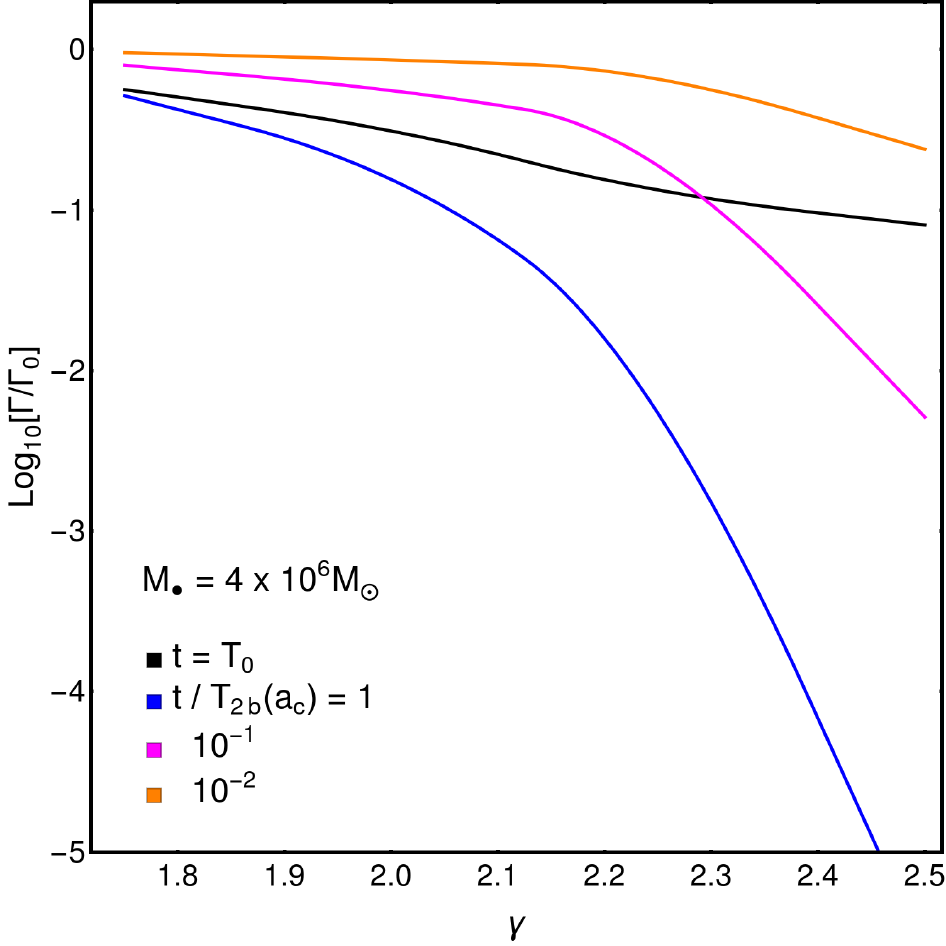}
    \caption{EMRI rate ratio $\Gamma/\Gamma_0$ as a function of density slopes $\gamma$ for a MW type NSC with $\Mbh = 4 \times 10^6 \Msun$ at various times in color. The black curve corresponds to the suitable reference timescale $T_0$ for comparison with a steady-state undepleted BH cusp. The rates (at $T_0$) are suppressed by a factor of $\simeq 3$ for $\gamma = 2$, while becoming an order of magnitude lower only for very steep cusps with $\gamma \gtrsim 2.4$.   }
    \label{fig_rate_ratio}
\end{figure}

\begin{figure*}
    \centering
   \includegraphics[width=0.425\textwidth]{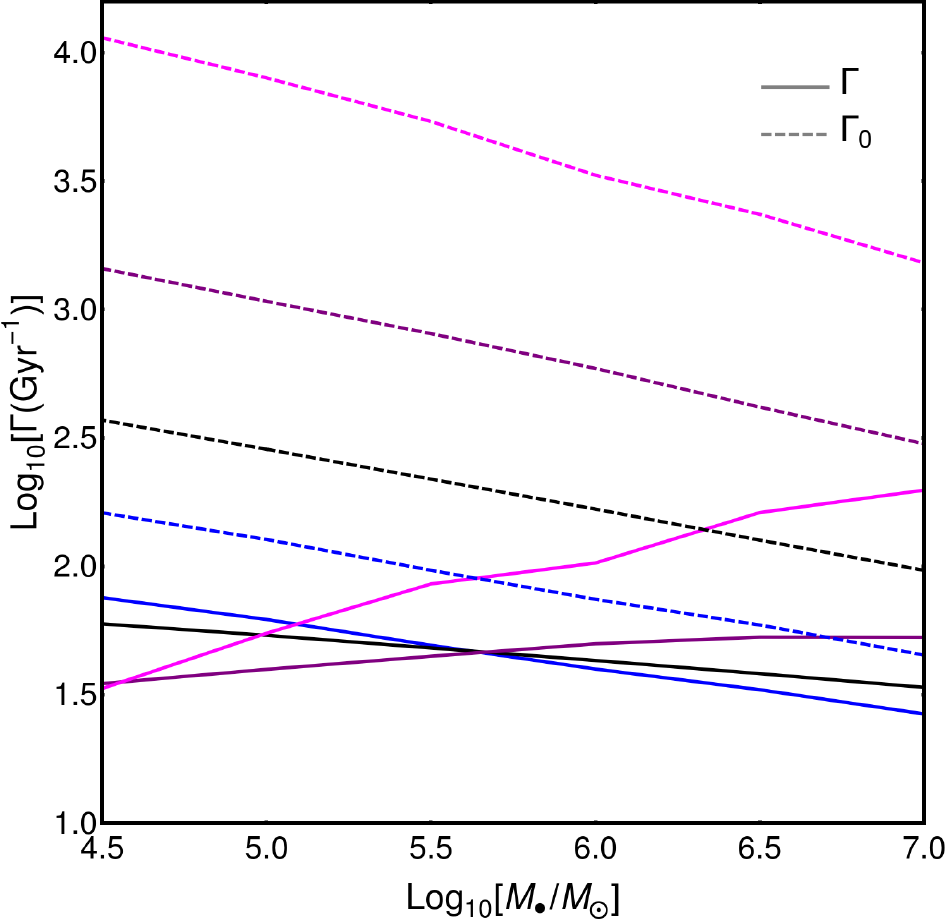}
    \hspace{1cm}
    \includegraphics[width=0.425\textwidth]{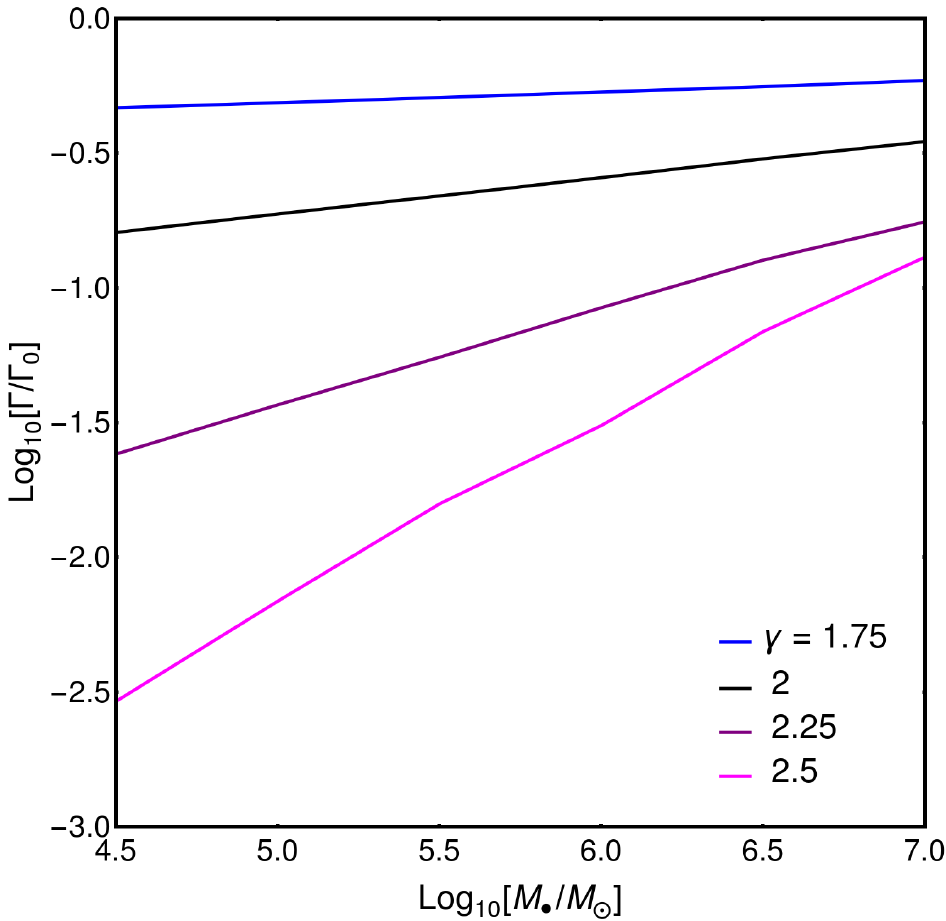}
    \caption{EMRI rates, $\Gamma$ (including ejections, solid curves) and $\Gamma_0$ (without ejections, dashed curves), in left panel, and their ratio $\Gamma/\Gamma_0$ in the right panel, are presented as a function of $\Mbh$ for various $\gamma$ (in color) at time $T_0$. Both lower MBH masses (low $\Mbh$) and steeper density slopes of clusters (high $\gamma$) support a higher number density of BHs in the inner regions, which can potentially increase the rates of both EMRIs and ejections. From the left panel, $\Gamma_0$ is higher for higher $\gamma$ and lower $\Mbh$, and roughly follows the classical relation $\Gamma_0 \propto \Mbh^{-1/4}$ found in many previous works without the inclusion of ejections. However, impact of ejections: (1) destroys these simplistic trends for EMRI rates $\Gamma$ that become an increasing function of MBH mass $\Mbh$ for steep profiles with $\gamma \gtrsim 2.25$, (2) caps the EMRI rates at a maximum $\Gamma \simeq 200 \,{\rm Gyr}^{-1}$ for the overall range of parameters explored here, while EMRI rates without ejections can attain high values of $\Gamma_0 \sim$ a few $\times 1000\,{\rm Gyr}^{-1}$. From the right panel, suppression is stronger for higher $\gamma$ and lower  $\Mbh$. For $\gamma = 2$, the rates are suppressed by a factor $\approx 3$ for all $\Mbh$, while the suppression becomes important by an order of magnitude for (a) $\Mbh \lesssim 10^6 \Msun$ for $\gamma = 2.25$, and (b) $\Mbh \lesssim 10^7 \Msun$ for $\gamma = 2.5$.  }
    \label{fig_rates_diff_para}
\end{figure*}

In this section, we discuss the results obtained from the numerical solution, while reporting the EMRI rates for a wide range of physically interesting parameters for NSCs. This includes MBH masses $\Mbh = 10^{4-7} \Msun$ relevant for the low-frequency GW detectors \citep{Babak2017}, and density slopes $\gamma = 1.75-2.5$ of BH cluster to check the impact of strong segregation \citep{Alexander_Hopman2009}. 
Further, we consider the cluster to host a single BH population with mass $m = 10 \Msun$, and choose a BH number fraction of $\fbh = 10^{-3} $ motivated by realistic stellar initial mass functions (IMFs; discussed below in more detail). This also ensures that the strong segregation regime is valid inside the influence radius $\rh$ of MBH \citep{Alexander_Hopman2009}. We employ the form of influence radius $ \rh = 2 {\rm pc}\sqrt{\Mbh/(4 \times 10^6 \Msun)}$, motivated by $\Mbh-\sigma$ relation \citep{Kormendy_Ho_2013}. So, the BH cusp has a total of $\Nf = 10^{-3} \Mbh/\Msun$ BHs within $\aref = 0.1 \rh$, considering most stars as solar-type with mass $\mstar = 1 \Msun$. Here, we present the time-dependent EMRI rates, $\Gamma$ (accounting for ejections) and $\Gamma_0$ (without the sink term induced by ejections) resulting from the numerical solution of FP equation~\ref{FP}. 

Most numerical works on strong segregation lead to relaxed BH cusps with $\gamma \sim 2$ \citep{Preto_Amaro_Seoane_2010}. Further, one can theoretically estimate $\gamma$ for various realistic stellar IMFs (\citealt{Salpeter_1955,Miller_Scalo_1979,Kroupa_2001}), by computing the associated relaxation coupling constant, $\Delta = 4 \fbh (m/\mstar)^2/(3+m/\mstar) $ which separates the strong-segregation regime ($\Delta \ll 1$) from that of weak segregation ($\Delta > 1$) \citep{Alexander_Hopman2009}. For a typical $m/\mstar \simeq 10$, these IMFs correspond to $\fbh \simeq$ $4 \times 10^{-4} - 10^{-3}$, leading to $\Delta \simeq 0.01-0.03$ \footnote{Here we assume that all stars $\mstar \geq 30 \Msun$, end up as stellar BHs with mass $m =10 \Msun$; while the remaining lighter stars either remain on Main Sequence branch or form compact objects of mass $\sim 1 \Msun$.} that implies $\gamma \simeq 2-2.2$ (see figure~4 of \citealt{Alexander_Hopman2009}). For $\gamma\geq2.5$, unrealistically low values of $\Delta \lesssim 10^{-3}$ are needed. Hence, the moderate slopes with $\gamma \approx 2-2.25$ can be deemed as the more realistic scenarios for the strong segregation regime.

First, we identify a critical semi-major axis $\aco$ from the numerical solution, within which EMRIs are expected to dominate over plunges. Its analytical counterpart $\ac$ is given in equation~\ref{ac}. Technically, $\aco$ corresponds to the initial semi-major axis which is connected to an EMRI (progenitor) orbit with \emph{final} $r_p = 4 R_s$ (at $a = \agw$) through a stream-line defined by the direction of the net-flux (see figure~\ref{fig_st_lines}). For this, we employ the final state reached (without the inclusion of ejections) at time $\Tbb(\ac)$ marking the end of simulations. So, only the fluxtubes originating at $a \leq \aco$ can channel BHs onto the inspiral orbits near MBH. Table~\ref{tbl_rates} gives both these lengthscales, $\aco$ and $\ac$, for some representative values of $\Mbh$ and $\gamma$. Evidently, $\aco \sim$ a few $\times \, 0.1-10$mpc, is always smaller than $\ac \sim$ a few $\times \, 10-100$mpc. This trend stands out especially for steeper cusps with high $\gamma$, because associated higher BH densities imply shorter 2B timescales, leading to a narrower GW loss cone defined by $r_p \leq \rpgw$ (see equation~\ref{rpgw}). 
While smaller physical values of $\aco$ are associated with smaller MBH masses and steeper BH cusps, the ratio $\aco/\rh$ turns out to be almost independent of $\Mbh$ throughout the full range of parameters, with $\aco/\rh \simeq \{0.018,\, 0.0094,\, 0.0032,\, 0.0007 \}$ for typical values $\gamma= \{1.75,\, 2,\, 2.25, \,2.5 \}$. Further, the values of critical semi-major axis $\aco$ recovered from our FP approach is compatible with those quoted in earlier Monte-Carlo approaches \citep{Raveh_2021}. 

We evaluate the instantaneous EMRI rates $\Gamma$ (and $\Gamma_0$ for simulations without considering ejections) from the numerical solution as the net advective flow in $a$-direction at the inner boundary $a = \agw$.
This presents a suitable method especially for the scenario including ejections, because the cusp is effectively depleted and ejections are not important inside $\agw$. Figure~\ref{fig_rate_evo} showcases the evolving EMRI rates for various $\Mbh$ and $\gamma$, where we also present the well-studied case of the Milky Way (MW)-type NSC (with $\Mbh =\Mmw =4 \times 10^6 \Msun$) in the left panel, enabling an easier comparison with earlier studies \citep{Hopman_Alexander_2005,Amaro_Seoane_Preto_2011,Raveh_2021} \footnote{For the case of $\gamma = 2$ cusp for MW-type NSC, the EMRI rate $\Gamma_0 \simeq 120~{\rm Gyr}^{-1}$ (without ejections) is compatible with previous works. With a slightly different parameters, \citet{Raveh_2021} 
evaluate steady state rates of $\sim 200~{\rm Gyr}^{-1}$ through a Monte-Carlo approach. This difference may also arise due to time-dependent effects, and early depletion of BHs at inner-most radii near $\agw$.}.
The impact of ejections is quite evident with lower EMRI rates $\Gamma$, in comparison to $\Gamma_0$ for the case without ejections. This rate suppression is strongest for the smallest $\Mbh$ and highest $\gamma$, though it builds up over time due to loss of BHs owing to ejections. The increasing magnitude of rate suppression with time is evident from figure~\ref{fig_rate_ratio}, which gives the ratio $\Gamma/\Gamma_0$ for $\Mbh=\Mmw$ at different values of time.

Figure~\ref{fig_rate_evo} shows the time evolution of EMRI rates $\Gamma$ and $\Gamma_0$, which is quite significant for steeper cusps. For $\gamma \leq 2.25$ with dominant diffusive fluxes $N/\Tbb \propto a^{2(2.25 -\gamma)}$ at larger $a$, there is only a moderate decay of rates over time  $T_0 = \Tbb(\aco)$ (gray vertical lines in the figure); see table~\ref{tbl_rates} for the typical values of $T_0$ for various $\Mbh$ and $\gamma$. This is because of the evolution of the initially isotropic system (uniform in $\scR$) towards the \citet{Cohn_Kulsrud_1978} logarithmic profile in the diffusion regime ($r_p \gtrsim \rpgw$) \footnote{However, one may note the steep decline of rates even for time shorter than $T_0$ for steep cusp $\gamma = 2.5$, which occurs due to depletion of sources at smaller $a$ that dominantly contribute towards diffusive fluxes.}. Afterward, rates, both $\Gamma$ and $\Gamma_0$, decay steeply due to the depletion of BHs (see figures~\ref{fig_Na_prof} and \ref{fig_Nq_prof}) owing to losses due to the capture by MBH and ejections driven by strong scatterings \footnote{We also check the evolution of the instantaneous ejection rates $\Nej$ of BHs with $a \leq \aco$, and compare it with that of EMRI rates $\Gamma$ in figure~\ref{fig_Nej_evo}.}. 
So, $T_0$ can serve as a reference time over which  BHs need to be replenished to channel a steady flux of EMRIs, and we quote the instantaneous values of $\Gamma$ and $\Gamma_0$ at this time (black points on the evolution curves) for a better comparison with steady rates reported in literature. From table~\ref{tbl_rates}, $T_0 \sim 10^{7-9}$yr for MW-type NSC, while for lower MBH masses this timescale can be shorter by upto 2 orders of magnitude. The replenishment of BHs (possibly leading to a steady-state) in real NSCs can occur as a result of dynamical friction arising from energy relaxation, that we do not consider here. Further, galactic nuclei may undergo episodes of star formation \citep{Levin_2003,Lu_2013,Aha+15,Aharon_Perets_2016,Schodel_2020}, 
that can also rejuvenate the supply of BHs. From figure~\ref{fig_rate_ratio} for the case of a MW-type NSC, rates $\Gamma$ (at the relevant time $T_0$) are suppressed by more than a factor of 3 for $\gamma \gtrsim 2$, while becoming an order of magnitude lower than $\Gamma_0$ for steepest cusps with $\gamma \gtrsim 2.4$.       

We report the EMRI rates over a wide range of $\Mbh$ and $\gamma$ in figure~\ref{fig_rates_diff_para} and table~\ref{tbl_rates} at the time $T_0$. The EMRI rates $\Gamma_0$, without ejections, are higher for steeper density profiles and lower MBH masses following roughly $\Gamma_0 \propto \Mbh^{-1/4}$ (for $\gamma \leq 2.25$), as obtained previously in the works without consideration of strong scatterings \footnote{This correlation arises from the proportionality $\Gamma_0 \propto 1/\TKep(\rh)$, and using $\Mbh-\sigma$ relation.} \citep{Hopman_Alexander_2005,Aharon_Perets_2016,Broggi_2022}. For realistic values of $\gamma \simeq 2 - 2.25$, $\Gamma_0 \simeq$ $ 100-1500~{\rm Gyr}^{-1}$ throughout the explored span of $\Mbh$, and the suppressed rates $\Gamma$ due to ejections, are lower by a factor of $\approx 3-30$ depending on $\Mbh$ and $\gamma$. As star clusters with high $\gamma$ and low $\Mbh$ are expected to have higher BH number densities $\propto \Mbh/\rh^3 \propto \Mbh^{-1/2}$, both the probability of ejections and EMRI formation tend to be higher for these systems. This effect levels off the suppressed rates to $\Gamma\simeq 30-50~{\rm Gyr}^{-1}$ for all $\Mbh$ and realistic segregated profiles with $\gamma \simeq 2-2.25$. The extremely steep cusps with $\gamma = 2.5$ retain $\Gamma \simeq30-200~{\rm Gyr}^{-1}$, although these are suppressed by a factor $\approx 10-300$ for all $\Mbh$. However, for unsegregated cusps with BW profile $\gamma = 1.75$, the rate suppression occurs upto a factor of $\simeq 2$ for all $\Mbh$. It is notable that usual EMRI rates $\Gamma_0$ span a large range of values with a factor $\approx 30-60$ for various $\gamma \in [1.75,2.5]$ for a given $\Mbh$. This expanse of values attained is more restricted for suppressed rates $\Gamma$, and is limited to a factor of $\approx 2-7$ for $\Mbh = 10^{5-7} \Msun$. Further, $\Gamma$ becomes an increasing function of $\Mbh$ for steeper profiles $\gamma \gtrsim 2.25$.

Figure~\ref{fig_rates_diff_para} (right panel) summarizes the level of suppression for all parameters. For a given $\Mbh$, the rate suppression due to ejections becomes increasingly important for steeper profiles with larger $\gamma$. This behavior is similar to that found by \cite{Teboul_2024} for TDE rates. For non-segregated BW cusps with $\gamma=1.75$, ejections suppress the EMRI rates by a factor of $\approx 2$. For segregated cusp with $\gamma = 2$, the suppression level slightly increases upto a factor of $\approx 3$. For the realistic and moderately steep profiles with $\gamma = 2.25$, the rates are suppressed by more than an order of magnitude for $\Mbh \lesssim 10^6 \Msun$. Highly steep cusps with $\gamma = 2.5$ have rates suppressed upto a factor $ \gtrsim 10$ for all $\Mbh$, while suppression gets important upto 2 orders of magnitude for $\Mbh \lesssim 10^5 \Msun$. These lower MBH masses otherwise have the highest unsuppressed rates $\Gamma_0$ and can channel EMRIs at LISA frequencies with the highest sensitivity \citep{Babak2017}, and are thus expected to dominate the future detection rates in the standard relaxation-driven scenario of EMRI formation. Thus, ejections driven by strong scatterings in strongly segregated cusps can have significant consequences for upcoming EMRI detections.   

\section{Summary and Discussion} 
\label{sec_discus}

Strong scatterings can significantly alter the loss cone dynamics around massive black holes in galactic nuclei \citep{Teboul_2024}. In the current study, we hereby test the relevance of strong scatterings for EMRI formation in a strongly segregated BH cusp lying in the inner regions of an NSC. We numerically solve a time-dependent two-dimensional Fokker-Planck equation in an effective energy-angular momentum space to track the evolution of BH distribution. This takes into account angular momentum relaxation driven by 2B scatterings, GW induced losses, and ejections triggered by strong scatterings. We also improve upon the previous treatment of orbit-averaged ejection rates, by considering depletion of inner cusp of strong scatterers due to GW loss cone (see equations~\ref{Pdotav_gen}-\ref{FejB} for exact expression, and equation~\ref{Pdotav_approx1} for a good approximation).

Employing the FP solution, we evaluate the time-dependent EMRI rates, that decrease with time due to depleting BHs (figure~\ref{fig_rate_evo}), as a result of the capture by central MBH and ejections. The resulting EMRI rates $\Gamma$ are lower than those $\Gamma_0$ without consideration of ejections, however the magnitude of rate suppression depends sensitively on the central MBH mass $\Mbh$ and steepness of BH cusp defined by density profile index $\gamma$. We explore a wide and physically interesting range of $\gamma=1.75-2.5$ and $\Mbh= 10^{4-7} \Msun$ for a single BH population of mass $10 \Msun$ with a fixed BH number fraction of $\fbh = 10^{-3}$ within $a \leq 0.1 \rh$. The choice of a compact BH distribution with a low $\fbh$ is motivated by the previous studies on EMRI formation in strongly segregated cusps \citep{Amaro_Seoane_Preto_2011,Raveh_2021}.

We find that the impact of strong scatterings limits the per-galaxy rate of EMRI formation to the highest value of $\Gamma\simeq 200~{\rm Gyr}^{-1}$ for all $\Mbh $ and $\gamma$ (Figure~\ref{fig_rates_diff_para} left panel), which otherwise can reach upto $\Gamma_0 \sim$ a few $\times 10^{3}~{\rm Gyr}^{-1}$ for steep BH cusps around low-mass MBHs. Contrary to the usual trend, suppressed rates $\Gamma$ become an increasing function of $\Mbh$ for steep cusps with $\gamma \gtrsim 2.25$. The level of suppression is highest for steepest cusps and lowest mass MBHs (Figure~\ref{fig_rates_diff_para} right panel). Weakly segregated cusps with Bahcall-Wolf $\gamma=7/4$ profile, display only moderate rate suppression upto a factor of 2. However, for strongly segregated BH cusps with $\gamma \gtrsim 2.25$, the rate suppression becomes significant upto an order of magnitude for $\Mbh \lesssim 10^6 \Msun$. For steepest cusps with $\gamma = 2.5$, rates are suppressed upto 2 orders of magnitude for small MBH masses of $\Mbh\lesssim 10^{5} \Msun$. Thus, the phenomenon of strong scatterings, which is not traditionally accounted for EMRI formation, can significantly suppress the EMRI rates in galactic nuclei. The high levels of rate suppression for steep BH cusps around low-mass MBHs, brings forth the significance of strong scatterings for future EMRI detections by LISA, which is expected to be maximally sensitive for these low-mass MBHs \citep{Babak2017}.

We note that our framework captures the time-dependent effects only partially due to the assumption of a non-evolving background field of scatterers (see for eg., \citealt{Vasiliev_17,Pan_Yang21,Broggi_2022} for a consistent time-dependent approach). This fixes in time both the: (1) the diffusion coefficient for $L$-relaxation owing to weak 2B scatterings, and (2) ejection probability rate due to strong scatterings, which would otherwise decay with time because of depletion of BHs. However, this effect of expected depletion might be naturally countered by replenishment of BHs due to energy relaxation (that is not considered here) leading to dynamical friction driven in-flow of BHs into these inner regions of NSC. Further, the growth of NSC owing to in-situ star formation episodes and/or mergers with star clusters sinking-in from larger distances, can contribute towards the replenishment. This may result into a quasi steady state of scatterers, similar to our assumption. Further, consideration of multiple BH populations may also impact the EMRI rates \citep{Aharon_Perets_2016}, by altering the rate of ejections \citep{Hen+69}. We defer these more detailed investigations to a future study. 

Our work brings out the significance of ejections driven by strong scatterings for EMRI formation in strongly segregated NSCs. However, the actual level of suppression implied for the future EMRI detection rates will depend sensitively on the structure and evolution of BH cusps in real NSCs and the lower end of the MBH mass function. The upcoming observational facilities, LISA and TianQin, necessitate more thorough future studies to evaluate realistic detection rates of these gravitational wave transients.

\begin{acknowledgments}
We would like to thank Odelia Teboul and Kartick C. Sarkar for helpful comments and discussions. KK gratefully acknowledges the support to access HPC at Technion, and computing facilities at Indian Institute of Technology, Kanpur.   

\end{acknowledgments}

\appendix

\section{Ejection rate driven by strong scatterings}
\label{app_ej_rate}

We consider a subject BH of mass $m$ moving with velocity $\vec{v}$ in a uniform-density sea of similar BHs with an isotropic distribution function $F$. The probability that the BH undergoes a velocity kick $\Delta \vec{v}$ per unit $\rmd^3 \Delta \vec{v}$ per unit time \citep{Agekyan_1959,Henon_1960}:
\beq 
R(\vec{v},\Delta \vec{v}) = \frac{8 \pi G^2 m^2}{(\Delta v)^5} \, \int_{v''}^{\infty} \rmd v' \, v' F(v')
\eeq 
where $v'' = \Delta v \, |1 + \vec{v} \centerdot \Delta \vec{v} /(\Delta v)^2|$ for equal-mass encounters considered here. 

The BH can get ejected from the system if its final speed exceeds that escape speed, i.e. the condition, $\mathcal{C} : |\vec{v} + \Delta \vec{v}| \geq \vesc$ , is satisfied.
This implies a probability of ejection per unit time, $\Pdot = \int_{\mathcal{C}} R(\vec{v},\Delta \vec{v}) \, \rmd^3 \Delta \vec{v} $, given explicitly as \citep{Henon_1960b,Goodman_1983}: 
\beq 
\Pdot = \frac{32 \pi^2 G^2 m^2 }{3 v (\vesc^2 - v^2)^2} \int_{\sqrt{\vesc^2 - v^2}}^{\vesc} \rmd v' \, v' \, F(v') \, (v'^2 + v^2 - \vesc^2)^{3/2}
\eeq 
The above integral is analytical for a near-Keplerian system and can be explicitly written as \citep{Teboul_2024}:
\beq 
\Pdot = P_0 \, a^2 \, v^{1+ 2 \gamma} \quad ; \quad \mbox{with} \quad P_0 = \frac{\Gamma(\gamma+1)}{\Gamma(\gamma+1/2)} \bigg(\frac{m}{\Mbh} \bigg)^2 \frac{\Nref}{\aref^3} \bigg( \frac{\aref}{2 G \Mbh} \bigg)^{\gamma}
\label{Pdot_fin}
\eeq 
for a power-law number profile $N(r) = \Nref (r/\aref)^{3-\gamma}$ of BH cusp, with reference BH number $\Nref = \fbh \Mbh/\mstar$ within $\aref$, as considered earlier in Section~\ref{sec_phy_setup}. This corresponds to a distribution function $F \propto \varepsilon^{\gamma - 3/2}$.   

Hence, the orbit-averaged ejection rate $\Pdotav$ for the BH on an orbit with elements $\{a ,r_p \}$ can be evaluated as $\Pdotav= (\pi \TKep(a))^{-1} \int \rmd r \, \Pdot v_r^{-1}$, with $v_r$ as the radial speed along the Keplerian orbit at given phase $r$. We approximate this integral by considering only the dominant contribution at the phase $r_0 = {\rm max}[r_p, \agw]$, which gives the following simplified expression:  
\beq 
\Pdotav \simeq  \frac{P_0 \, a^2}{\pi \TKep(a)} \, r_0 \, \bigg( \frac{2 G \Mbh}{r_0} - \frac{G \Mbh}{a} \bigg)^{\gamma}
\label{Pdotav_approx2}
\eeq 
Using equation~\ref{Tbb}, the above expression can be posed in an alternative form presented in equation~\ref{Pdotav_approx1}. 

The above expression for $\Pdotav$ is accurate to a few tens of per cent to the actual integral in the expression for $\Pdotav$, that is derived below. But, first we need to model the distribution of BHs in the depleted cusp within $\agw$. 

\subsection{Depleted cusp}

We assume that initial BH number profile $N(a) = \Nref (a/\aref)^{3-\gamma}$ continues till $a \geq \agw$, while orbital evolution for $a\lesssim \agw$ is dominated by GW emission even for circular orbits. Hence, depletion due to capture of stellar BHs by central MBH becomes important inside $\agw$, and the initial density profile with power-law index $\gamma$ can not continue inside this critical distance.  

Since the orbits with $a \leq \agw$ correspond to EMRI progenitors, these are sourced by diffusive fluxes in the outer regions with semi-major axis $ \agw <a' < \ac$. We denote the depleted profile of scatterers $\Nd(a)$ for $a\leq \agw$, which can be estimated in the steady state limit \citep{Sari_Fragione_2019,KaurFP_24} as $ {\Nd(a)}/{\Tgw(a)} = {N(a')}/({\lnLo \Tbb(a')})$. Here $\lnLo \simeq \log{(a'/\rpgw(a'))} \simeq 10$ is the logarithmic size of GW loss cone \footnote{Though $\lnLo$ depends on the associated semi-major axis $a'$, we treat it as effectively a constant, and choose $\lnLo \simeq 10$ which is valid at $a' = \ac$.}, and $a' > \agw$ is the initial semi-major axis that channels the BHs at $a = \rpgw(a') < \agw$, where the loss cone boundary $\rpgw$ is given by equation~\ref{rpgw}. The inherent assumption in this approach is that BH orbits predominantly diffuse in $r_p$ due to 2B scatterings for $r_p > \rpgw$ , while for $r_p < \rpgw$ advection in $a$ occurs due to GW emission. This gives a power-law form of depleted cusp: 
\beq
 \Nd(a) = \Nrefd \bigg(\frac{a}{\agw}\bigg)^{3-\gamd} \qquad
 \mbox{where  } \quad \Nrefd = \frac{N(\agw)}{\lnLo} = \frac{\Nref}{\lnLo} \bigg(\frac{\agw}{\aref}\bigg)^{3-\gamma} \quad 
\mbox{and }\quad \gamd = \frac{33 - 16 \gamma}{4 (3 - \gamma)} \,. 
\label{Nd_depleted}
\eeq 
The depleted profile is much shallower with a power-law index $\gamd$ which is always a decreasing function of $\gamma$. 
For $\gamma \in [1.75,2.5]$, $\gamd \in [ - 3.5  , 1]$ where the change in sign occurs at $\gamma \simeq 2.06$. 

We assign a function $N'(a) = \Nref' (a/\aref')^{3-\gamma'}$ to the overall BH number profile, with piecewise functions:
\beq 
\{ \Nref' , \aref' ,\gamma'  \} = \begin{cases} 
   &  \{\Nref, \aref, \gamma  \} \qquad \,, \quad a \geq \agw      \\
    & \{\Nrefd, \agw, \gamd  \} \quad, \quad a < \agw
\end{cases} 
\label{N_overall}
\eeq 

\subsection{Orbit-averaged ejection rate} 

Here we formulate an exact way to evaluate orbit-averaged ejection rate $\Pdotav$ of a BH with orbital elements $\{a,r_p \}$. For this, we also need a general local ejection rate $\Pdot$, valid for the entire test BH orbit as it traces both undepleted and depleted portions of the cusp, which is defined as: 
\beq 
\begin{split}
& \Pdot = P_0' a^2 v^{1+2 \gamma'}  \quad \mbox{with piece-wise function:} \\
& P_0' = \begin{cases} 
      &  P_0 = \displaystyle{ \frac{\Gamma(\gamma+1)}{\Gamma(\gamma+1/2)} \bigg(\frac{m}{\Mbh} \bigg)^2 \frac{\Nref}{\aref^3} \bigg( \frac{\aref}{2 G \Mbh} \bigg)^{\gamma} \quad \mbox{for} \quad a \geq \agw }        \\[1em]
      &  P_{0{\rm d}} = \displaystyle{  \bigg(\frac{m}{\Mbh} \bigg)^2 \frac{\Nrefd}{\agw^3} \bigg( \frac{\agw}{2 G \Mbh} \bigg)^{\gamd} } \qquad \mbox{for} \quad a < \agw \\
\end{cases}
\end{split}
\label{Pdot_gen}
\eeq 
as implied by equation~\ref{Pdot_fin} and \ref{N_overall}. 

We can then orbit-average this local ejection rate over true anomaly $f$ as:
\beq 
\begin{split}
&\Pdotav = \frac{1}{2 \pi \TKep(a) \sqrt{G \Mbh a (1-e^2)}} \int_0^{2\pi} \rmd f \, r^2 \Pdot \, 
\,  =\, \frac{P_0' a^3 (G \Mbh/a)^{\gamma'}}{ 2 \pi \TKep(a)\sqrt{1-e^2}} \int_0^{2\pi} \rmd f \, r'^2 \, v'^{1+2 \gamma'} \\[1ex]
 & \mbox{where } \quad r' = \frac{1-e^2}{1+ e \cos{f}} \qquad, 
  \qquad  v' = \sqrt{\frac{2}{r'}-1} = \sqrt{\frac{1+e^2 +2 e \cos{f}}{1-e^2}} \,. 
\end{split}
\label{Pdotav_gen}
\eeq 

This leads to the final ejection term $\Fej = \Pdotav \Tbb(\aref)$ (defined earlier in equation~\ref{FP}) for the two cases, (a) $r_p \geq \agw$ where whole orbit remains in undepleted cusp, and (b) $r_p < \agw$ where a portion of the orbit penetrates into depleted cusp inside $\agw$.   

\bigskip

\noindent
{\bf Case a.} $r_p \geq \agw$ 

\smallskip 

\noindent
Since the orbit remains only in the undepleted cusp defined by density index $\gamma$, we have the following form of ejection term: 
\beq 
\Fej = \frac{\Gamma(\gamma+1)}{\Gamma(\gamma+1/2)} \frac{3\sqrt{2} \pi}{2^{\gamma+5}  C \lnL} \bigg( \frac{\aref}{a} \bigg)^{\gamma-3/2} \frac{\Int(e,\gamma,0,2\pi)}{(1-e^2)^{\gamma-1}} 
\label{FejA}
\eeq 
where the function $\Int$ is defined as: 
\beq 
\Int(e,\gamma,f_1,f_2) = \int_{f_1}^{f_2} \rmd f \, \frac{(1+e^2+2 e \cos{f})^{\gamma+1/2}}{(1+e \cos{f})^2} 
\label{Int}
\eeq 

\medskip

\noindent
{\bf Case b.} $r_p < \agw$

The BH orbit is inside the depleted cusp for $f \in [-\fgw,\fgw]$, while it remains in the undepleted cusp with $r > \agw$ for the remaining orbital phases, where
\beq 
\fgw = \cos^{-1}{\bigg( \frac{a(1-e^2)}{ \agw \, e} - \frac{1}{e} \bigg)} \,.
\eeq 
Hence, for this case the ejection term $\Fej = \Feju + \Fejd$, where $\Feju$ and $\Fejd$ arise due to the time spent in undepleted and depleted cusp respectively, and are given as: 
\beq 
\begin{split} 
& \Feju = \frac{\Gamma(\gamma+1)}{\Gamma(\gamma+1/2)} \frac{3\sqrt{2} \pi}{2^{\gamma+4}  C \lnL} \bigg( \frac{\aref}{a} \bigg)^{\gamma-3/2} \frac{\Int(e,\gamma,\fgw,\pi)}{(1-e^2)^{\gamma-1}} \,, \\
& \Fejd = \frac{3\sqrt{2} \pi}{2^{\gamd+4}  C \lnL \lnLo} \bigg(\frac{\aref}{\agw} \bigg)^{\gamma-\gamd} \bigg(\frac{a}{\aref} \bigg)^{3/2-\gamd} \frac{\Int(e,\gamd,0,\fgw)}{(1-e^2)^{\gamd-1}} \,.
\end{split} 
\label{FejB}
\eeq

\FloatBarrier

\section{Additional Figures} 

\begin{figure*}
    \centering
    \includegraphics[width=0.4\textwidth]{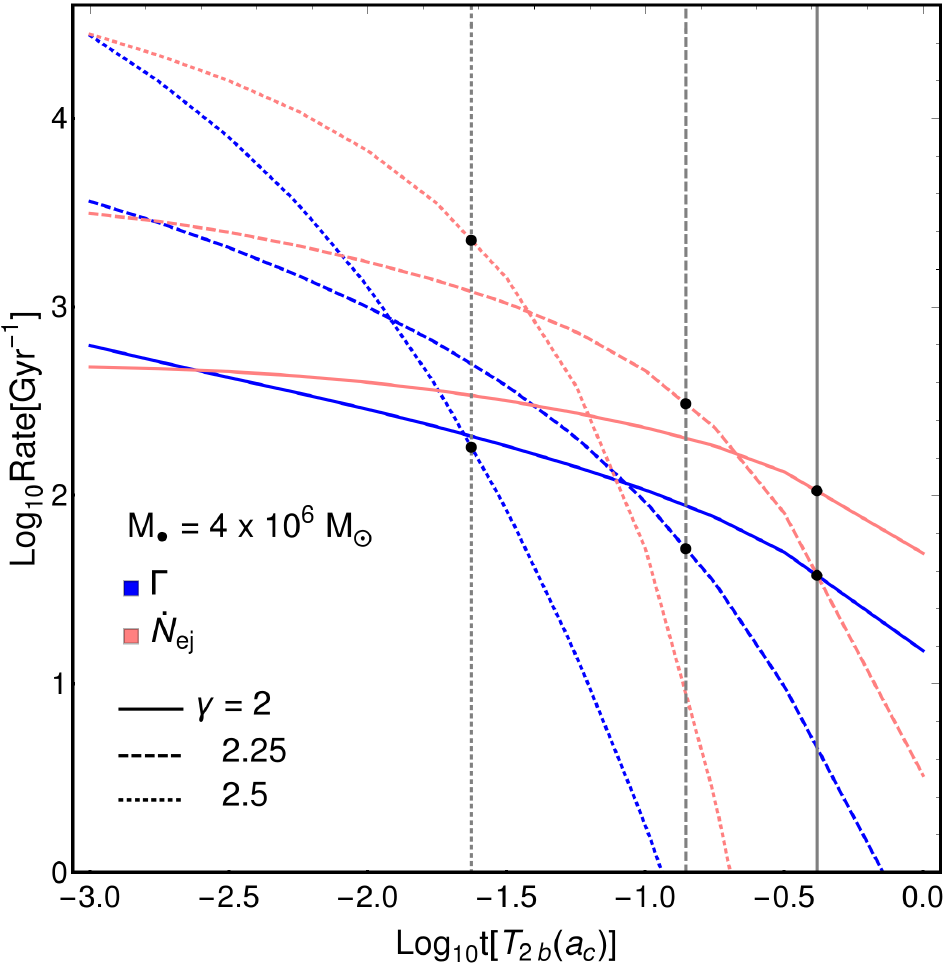}
    \hspace{1cm}
    \includegraphics[width=0.4\textwidth]{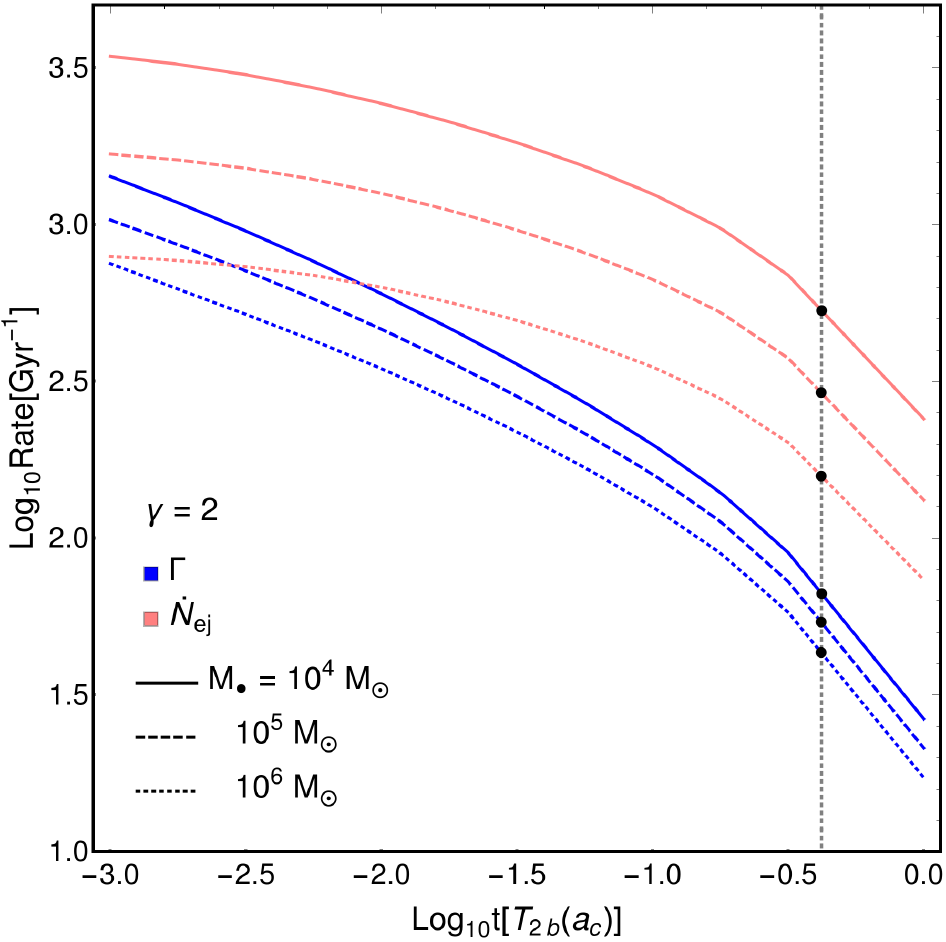}
    \caption{Evolution of EMRI rates $\Gamma$ (including strong scatterings; in blue) and ejection rates $\Nej$ (in pink) is shown as a function of time for [Left panel] for various $\gamma = 2 - 2.5$ and fixed $\Mbh = 4 \times 10^6 \Msun$, and [Right panel] different $\Mbh = 10^{4-6} \Msun$ and fixed $\gamma = 2$. The ejection rate $\Nej$ is computed for BHs with $a \leq \aco$, that are relevant for EMRI formation. As earlier, we depict the rates at time $T_0 = \Tbb(\aco)$ (vertical gray lines) as black points.  
    }
    \label{fig_Nej_evo}
\end{figure*}

\begin{figure*}
    \centering
    \includegraphics[width=1.\textwidth]{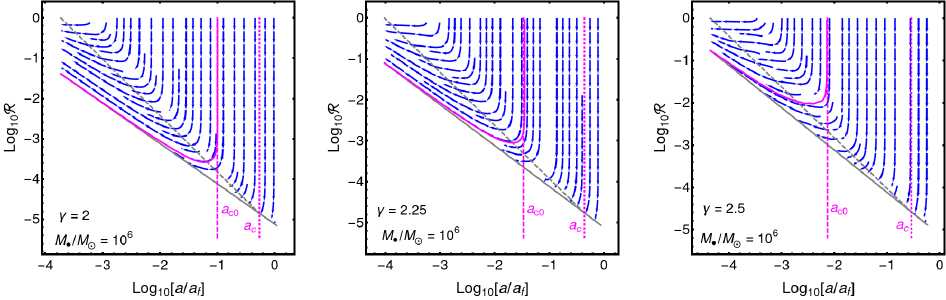}
    \caption{Evaluation of the critical semi-major axis $\aco$. Here, we show the streamlines (in blue) in $\{a, \scR \}$-plane indicating the net-flow of BHs at the end of simulations ($t = \Tbb(\ac)$) without strong scatterings. Different panels correspond to $\gamma = \{2, 2.25, 2.5 \}$ while fixing $\Mbh = 10^6 \Msun$. Evidently, the analytical estimate $\ac$ (dotted magenta line) of the critical semi-major axis (equation~\ref{ac}) is not completely accurate and stream-lines initiating at  $a \in (\aco,\, \ac)$ end up into plunging orbits inside the capture radius (solid gray line). Hence, $\aco \ll \ac$ is a better estimate of the critical semi-major axis, as most stream-lines inside with initial $a \leq \aco$ lead to inspiral orbits that eventually evolve into EMRIs. As $\gamma$ increases towards right, $\aco/\aref$ becomes smaller (and its departure from $\ac/\aref$ is higher) due to narrowing GW loss cone (region under the dashed gray line defined by $r_p \leq \rpgw$; equation~\ref{rpgw}). However, $\aco/\aref$ is almost independent of $\Mbh$, with $\aco/\aref = \{ 0.094,\, 0.032,\, 0.007 \}$ for $\gamma = \{ 2,\, 2.25,\,2.5 \}$.       }
    \label{fig_st_lines}
\end{figure*}

\begin{figure*} 
\centering
    \includegraphics[width=0.9\textwidth]{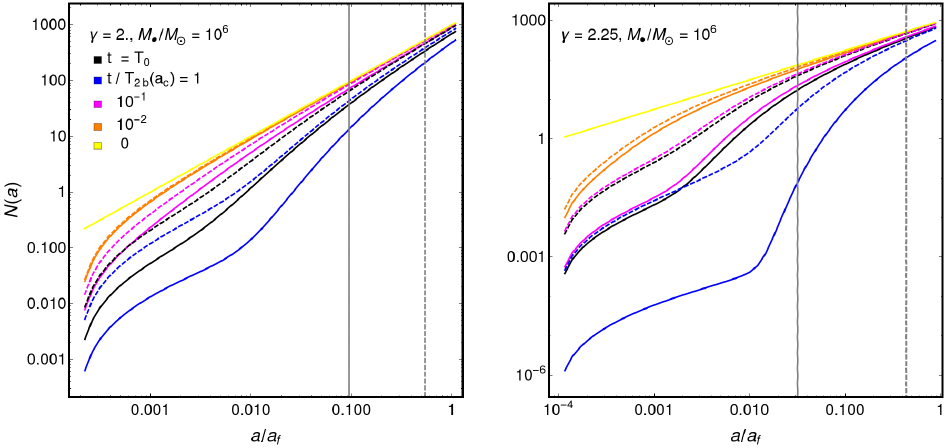}
    \caption{Evolution of BH number profiles $N(a)$, defined as number of BHs with semi-major axis $< a$, with time (in color). The solid (dashed) curves indicate the solution with (without) strong scatterings. The two panels correspond to $\gamma = 2,\, 2.25$ for fixed $\Mbh = 10^6 \Msun$. The black curves depict the solutions at the reference time $T_0$; while the initial power-law profile is depicted in yellow. The depletion of BHs due to ejections triggered by strong scatterings is evidently higher in the inner regions of the cusp (smaller $a$), and for clusters with higher $\gamma$. The kink in these curves approximately corresponds to the BH depletion due to the capture by central MBH, satisfying $\log{[1/\scR_{\rm cap}(a)]} \Tbb(a) \approx t$. The solid gray line corresponds to the critical semi-major axis $\aco$, while dashed gray line to its analytical counterpart $\ac$.  }
    \label{fig_Na_prof}
\end{figure*}

\begin{figure*} 
\centering
    \includegraphics[width=0.9\textwidth]{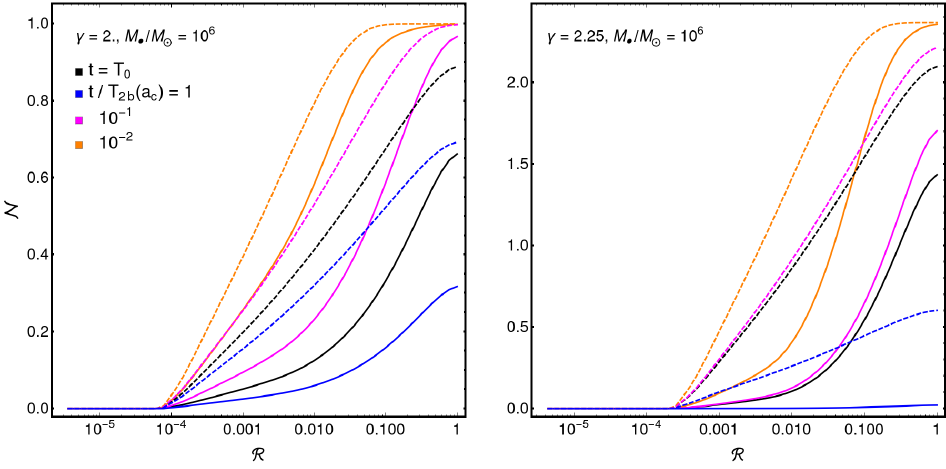}
    \caption{Evolution of 2D density $\scN$ of BHs, as function of $\scR$ at fixed $a = \aco$, with time in color. As earlier, solid (dashed) curves represent the solution with (without) strong scatterings, and black curves indicate the solution at time $T_0$. The panels are for $\gamma = 2, \, 2.25$ and fixed $\Mbh = 10^6 \Msun$. Evidently, the solutions without strong scatterings follow the expected logarithmic profile in $\scR$, while solutions, including this effect, is more depleted for low $\scR$ corresponding to higher eccentricity orbits, that are more prone to ejections. The impact of strong scatterings is higher for the cluster with higher $\gamma$ (right panel). }
    \label{fig_Nq_prof}
\end{figure*}

\FloatBarrier

\bibliography{main}{} 

\begin{thebibliography}{}
\expandafter\ifx\csname natexlab\endcsname\relax\def\natexlab#1{#1}\fi
\providecommand{\url}[1]{\href{#1}{#1}}
\providecommand{\dodoi}[1]{doi:~\href{http://doi.org/#1}{\nolinkurl{#1}}}
\providecommand{\doeprint}[1]{\href{http://ascl.net/#1}{\nolinkurl{http://ascl.net/#1}}}
\providecommand{\doarXiv}[1]{\href{https://arxiv.org/abs/#1}{\nolinkurl{https://arxiv.org/abs/#1}}}

\bibitem[{{Agekyan}(1959)}]{Agekyan_1959}
{Agekyan}, T.~A. 1959, \sovast, 3, 46

\bibitem[{{Aharon} \& {Perets}(2015)}]{Aha+15}
{Aharon}, D., \& {Perets}, H.~B. 2015, \apj, 799, 185, \dodoi{10.1088/0004-637X/799/2/185}

\bibitem[{{Aharon} \& {Perets}(2016)}]{Aharon_Perets_2016}
---. 2016, \apjl, 830, L1, \dodoi{10.3847/2041-8205/830/1/L1}

\bibitem[{{Alexander}(2017{\natexlab{a}})}]{Alexander_2017}
{Alexander}, T. 2017{\natexlab{a}}, \araa, 55, 17, \dodoi{10.1146/annurev-astro-091916-055306}

\bibitem[{{Alexander}(2017{\natexlab{b}})}]{Alexander_2017RR}
{Alexander}, T. 2017{\natexlab{b}}, in Journal of Physics Conference Series, Vol. 840, Journal of Physics Conference Series, 012019, \dodoi{10.1088/1742-6596/840/1/012019}

\bibitem[{{Alexander} \& {Hopman}(2009)}]{Alexander_Hopman2009}
{Alexander}, T., \& {Hopman}, C. 2009, \apj, 697, 1861, \dodoi{10.1088/0004-637X/697/2/1861}

\bibitem[{{Amaro-Seoane}(2018)}]{Amaro-Seoane_2018}
{Amaro-Seoane}, P. 2018, Living Reviews in Relativity, 21, 4, \dodoi{10.1007/s41114-018-0013-8}

\bibitem[{{Amaro-Seoane} {et~al.}(2007){Amaro-Seoane}, {Gair}, {Freitag}, {Miller}, {Mandel}, {Cutler}, \& {Babak}}]{Amaro-Seoane_2007}
{Amaro-Seoane}, P., {Gair}, J.~R., {Freitag}, M., {et~al.} 2007, Classical and Quantum Gravity, 24, R113, \dodoi{10.1088/0264-9381/24/17/R01}

\bibitem[{{Amaro-Seoane} \& {Preto}(2011)}]{Amaro_Seoane_Preto_2011}
{Amaro-Seoane}, P., \& {Preto}, M. 2011, Classical and Quantum Gravity, 28, 094017, \dodoi{10.1088/0264-9381/28/9/094017}

\bibitem[{{Ashurov}(2004)}]{Ashurov2004}
{Ashurov}, A.~E. 2004, \aj, 127, 2154, \dodoi{10.1086/382840}

\bibitem[{{Babak} {et~al.}(2017){Babak}, {Gair}, {Sesana}, {Barausse}, {Sopuerta}, {Berry}, {Berti}, {Amaro-Seoane}, {Petiteau}, \& {Klein}}]{Babak2017}
{Babak}, S., {Gair}, J., {Sesana}, A., {et~al.} 2017, \prd, 95, 103012, \dodoi{10.1103/PhysRevD.95.103012}

\bibitem[{{Bahcall} \& {Wolf}(1977)}]{Bahcall_Wolf_1977}
{Bahcall}, J.~N., \& {Wolf}, R.~A. 1977, \apj, 216, 883, \dodoi{10.1086/155534}

\bibitem[{{Bar-Or} \& {Alexander}(2016)}]{Bar-Or_2016}
{Bar-Or}, B., \& {Alexander}, T. 2016, \apj, 820, 129, \dodoi{10.3847/0004-637X/820/2/129}

\bibitem[{{Binney} \& {Tremaine}(1987)}]{Binney_Tremaine}
{Binney}, J., \& {Tremaine}, S. 1987, {Galactic dynamics}

\bibitem[{{Bode} \& {Wegg}(2014)}]{Bode_2014}
{Bode}, J.~N., \& {Wegg}, C. 2014, \mnras, 438, 573, \dodoi{10.1093/mnras/stt2227}

\bibitem[{{Bortolas} \& {Mapelli}(2019)}]{Bortolas2019}
{Bortolas}, E., \& {Mapelli}, M. 2019, \mnras, 485, 2125, \dodoi{10.1093/mnras/stz440}

\bibitem[{{Broggi} {et~al.}(2022){Broggi}, {Bortolas}, {Bonetti}, {Sesana}, \& {Dotti}}]{Broggi_2022}
{Broggi}, L., {Bortolas}, E., {Bonetti}, M., {Sesana}, A., \& {Dotti}, M. 2022, \mnras, 514, 3270, \dodoi{10.1093/mnras/stac1453}

\bibitem[{{Cohn} \& {Kulsrud}(1978)}]{Cohn_Kulsrud_1978}
{Cohn}, H., \& {Kulsrud}, R.~M. 1978, \apj, 226, 1087, \dodoi{10.1086/156685}

\bibitem[{{eLISA Consortium} {et~al.}(2013){eLISA Consortium}, {Amaro Seoane}, {Aoudia}, {Audley}, {Auger}, {Babak}, {Baker}, {Barausse}, {Barke}, {Bassan}, {Beckmann}, {Benacquista}, {Bender}, {Berti}, {Bin{\'e}truy}, {Bogenstahl}, {Bonvin}, {Bortoluzzi}, {Brause}, {Brossard}, {Buchman}, {Bykov}, {Camp}, {Caprini}, {Cavalleri}, {Cerdonio}, {Ciani}, {Colpi}, {Congedo}, {Conklin}, {Cornish}, {Danzmann}, {de Vine}, {DeBra}, {Dewi Freitag}, {Di Fiore}, {Diaz Aguilo}, {Diepholz}, {Dolesi}, {Dotti}, {Fern{\'a}ndez Barranco}, {Ferraioli}, {Ferroni}, {Finetti}, {Fitzsimons}, {Gair}, {Galeazzi}, {Garcia}, {Gerberding}, {Gesa}, {Giardini}, {Gibert}, {Grimani}, {Groot}, {Guzman Cervantes}, {Haiman}, {Halloin}, {Heinzel}, {Hewitson}, {Hogan}, {Holz}, {Hornstrup}, {Hoyland}, {Hoyle}, {Hueller}, {Hughes}, {Jetzer}, {Kalogera}, {Karnesis}, {Kilic}, {Killow}, {Klipstein}, {Kochkina}, {Korsakova}, {Krolak}, {Larson}, {Lieser}, {Littenberg}, {Livas}, {Lloro}, {Mance}, {Madau}, {Maghami}, {Mahrdt}, {Marsh}, {Mateos}, {Mayer},
  {McClelland}, {McKenzie}, {McWilliams}, {Merkowitz}, {Miller}, {Mitryk}, {Moerschell}, {Mohanty}, {Monsky}, {Mueller}, {M{\"u}ller}, {Nelemans}, {Nicolodi}, {Nissanke}, {Nofrarias}, {Numata}, {Ohme}, {Otto}, {Perreur-Lloyd}, {Petiteau}, {Phinney}, {Plagnol}, {Pollack}, {Porter}, {Prat}, {Preston}, {Prince}, {Reiche}, {Richstone}, {Robertson}, {Rossi}, {Rosswog}, {Rubbo}, {Ruiter}, {Sanjuan}, {Sathyaprakash}, {Schlamminger}, {Schutz}, {Sch{\"u}tze}, {Sesana}, {Shaddock}, {Shah}, {Sheard}, {Sopuerta}, {Spector}, {Spero}, {Stanga}, {Stebbins}, {Stede}, {Steier}, {Sumner}, {Sun}, {Sutton}, {Tanaka}, {Tanner}, {Thorpe}, {Tr{\"o}bs}, {Tinto}, {Tu}, {Vallisneri}, {Vetrugno}, {Vitale}, {Volonteri}, {Wand}, {Wang}, {Wanner}, {Ward}, {Ware}, {Wass}, {Weber}, {Yu}, {Yunes}, \& {Zweifel}}]{eLISA_Consort2013}
{eLISA Consortium}, {Amaro Seoane}, P., {Aoudia}, S., {et~al.} 2013, arXiv e-prints, arXiv:1305.5720.
\newblock \doarXiv{1305.5720}

\bibitem[{{Freitag}(2001)}]{Freitag_2001}
{Freitag}, M. 2001, Classical and Quantum Gravity, 18, 4033, \dodoi{10.1088/0264-9381/18/19/309}

\bibitem[{{Ghez} {et~al.}(2008){Ghez}, {Salim}, {Weinberg}, {Lu}, {Do}, {Dunn}, {Matthews}, {Morris}, {Yelda}, {Becklin}, {Kremenek}, {Milosavljevic}, \& {Naiman}}]{Ghez_2008}
{Ghez}, A.~M., {Salim}, S., {Weinberg}, N.~N., {et~al.} 2008, \apj, 689, 1044, \dodoi{10.1086/592738}

\bibitem[{{Gillessen} {et~al.}(2009){Gillessen}, {Eisenhauer}, {Trippe}, {Alexander}, {Genzel}, {Martins}, \& {Ott}}]{Gillessen_2009}
{Gillessen}, S., {Eisenhauer}, F., {Trippe}, S., {et~al.} 2009, \apj, 692, 1075, \dodoi{10.1088/0004-637X/692/2/1075}

\bibitem[{{Goodman}(1983)}]{Goodman_1983}
{Goodman}, J. 1983, \apj, 270, 700, \dodoi{10.1086/161161}

\bibitem[{{H{\'e}non}(1960{\natexlab{a}})}]{Henon_1960}
{H{\'e}non}, M. 1960{\natexlab{a}}, Annales d'Astrophysique, 23, 467

\bibitem[{{H{\'e}non}(1960{\natexlab{b}})}]{Henon_1960b}
---. 1960{\natexlab{b}}, Annales d'Astrophysique, 23, 668

\bibitem[{{Henon}(1969)}]{Hen+69}
{Henon}, M. 1969, \aap, 2, 151

\bibitem[{{H{\'e}non}(2011)}]{Henon11}
{H{\'e}non}, M. 2011, arXiv e-prints, arXiv:1103.3498, \dodoi{10.48550/arXiv.1103.3498}

\bibitem[{{Hils} \& {Bender}(1995)}]{Hils_1995}
{Hils}, D., \& {Bender}, P.~L. 1995, \apjl, 445, L7, \dodoi{10.1086/187876}

\bibitem[{{Hopman} \& {Alexander}(2005)}]{Hopman_Alexander_2005}
{Hopman}, C., \& {Alexander}, T. 2005, \apj, 629, 362, \dodoi{10.1086/431475}

\bibitem[{{Kaur} {et~al.}(2024){Kaur}, {Rom}, \& {Sari}}]{KaurFP_24}
{Kaur}, K., {Rom}, B., \& {Sari}, R. 2024, arXiv e-prints, arXiv:2406.07627, \dodoi{10.48550/arXiv.2406.07627}

\bibitem[{{Kormendy} \& {Ho}(2013)}]{Kormendy_Ho_2013}
{Kormendy}, J., \& {Ho}, L.~C. 2013, \araa, 51, 511, \dodoi{10.1146/annurev-astro-082708-101811}

\bibitem[{Kroupa(2001)}]{Kroupa_2001}
Kroupa, P. 2001, Monthly Notices of the Royal Astronomical Society, 322, 231, \dodoi{10.1046/j.1365-8711.2001.04022.x}

\bibitem[{{Levin}(2007)}]{Levin_2007}
{Levin}, Y. 2007, \mnras, 374, 515, \dodoi{10.1111/j.1365-2966.2006.11155.x}

\bibitem[{{Levin} \& {Beloborodov}(2003)}]{Levin_2003}
{Levin}, Y., \& {Beloborodov}, A.~M. 2003, \apjl, 590, L33, \dodoi{10.1086/376675}

\bibitem[{{Lightman} \& {Shapiro}(1977)}]{Lightman_Shapiro1977}
{Lightman}, A.~P., \& {Shapiro}, S.~L. 1977, \apj, 211, 244, \dodoi{10.1086/154925}

\bibitem[{{Lin} \& {Tremaine}(1980)}]{Lin_Tremaine_1980}
{Lin}, D.~N.~C., \& {Tremaine}, S. 1980, \apj, 242, 789, \dodoi{10.1086/158513}

\bibitem[{{Linial} \& {Sari}(2022)}]{Linial_Sari_22}
{Linial}, I., \& {Sari}, R. 2022.
\newblock \doarXiv{2206.14817}

\bibitem[{{Lu} {et~al.}(2013){Lu}, {Do}, {Ghez}, {Morris}, {Yelda}, \& {Matthews}}]{Lu_2013}
{Lu}, J.~R., {Do}, T., {Ghez}, A.~M., {et~al.} 2013, \apj, 764, 155, \dodoi{10.1088/0004-637X/764/2/155}

\bibitem[{{Mei} {et~al.}(2021){Mei}, {Bai}, {Bao}, {Barausse}, {Cai}, {Canuto}, {Cao}, {Chen}, {Chen}, {Ding}, {Duan}, {Fan}, {Feng}, {Fu}, {Gao}, {Gao}, {Gong}, {Gou}, {Gu}, {Gu}, {He}, {Hendry}, {Hong}, {Hu}, {Hu}, {Hu}, {Huang}, {Huang}, {Jiang}, {Jiang}, {Jiang}, {Jiang}, {Jin}, {Korol}, {Li}, {Li}, {Li}, {Li}, {Li}, {Li}, {Li}, {Li}, {Li}, {Liang}, {Liang}, {Liao}, {Liu}, {Liu}, {Liu}, {Liu}, {Liu}, {Liu}, {Liu}, {Lu}, {Lu}, {Lu}, {Luo}, {Luo}, {Milyukov}, {Ming}, {Pi}, {Qin}, {Qu}, {Sesana}, {Shao}, {Shi}, {Su}, {Tan}, {Tan}, {Tan}, {Tu}, {Wang}, {Wang}, {Wang}, {Wang}, {Wang}, {Wang}, {Wang}, {Wang}, {Wang}, {Wang}, {Wang}, {Wei}, {Wu}, {Xiao}, {Xu}, {Xue}, {Yang}, {Yang}, {Yang}, {Yang}, {Ye}, {Yeh}, {Yu}, {Zhai}, {Zhang}, {Zhang}, {Zhang}, {Zhang}, {Zhang}, {Zhang}, {Zhang}, {Zhou}, {Zhou}, {Zhou}, {Zhu}, {Zi}, \& {Luo}}]{Mei_2021}
{Mei}, J., {Bai}, Y.-Z., {Bao}, J., {et~al.} 2021, Progress of Theoretical and Experimental Physics, 2021, 05A107, \dodoi{10.1093/ptep/ptaa114}

\bibitem[{{Merritt}(2005)}]{Merritt05}
{Merritt}, D. 2005, \apj, 628, 673, \dodoi{10.1086/429398}

\bibitem[{{Merritt}(2013)}]{Merritt_2013}
---. 2013, {Dynamics and Evolution of Galactic Nuclei}

\bibitem[{{Merritt}(2015)}]{Merritt_2015}
---. 2015, \apj, 814, 57, \dodoi{10.1088/0004-637X/814/1/57}

\bibitem[{{Merritt} {et~al.}(2011){Merritt}, {Alexander}, {Mikkola}, \& {Will}}]{Merritt_2011}
{Merritt}, D., {Alexander}, T., {Mikkola}, S., \& {Will}, C.~M. 2011, \prd, 84, 044024, \dodoi{10.1103/PhysRevD.84.044024}

\bibitem[{{Merritt} {et~al.}(2007){Merritt}, {Berczik}, \& {Laun}}]{Merritt07}
{Merritt}, D., {Berczik}, P., \& {Laun}, F. 2007, \aj, 133, 553, \dodoi{10.1086/510294}

\bibitem[{{Miller} \& {Scalo}(1979)}]{Miller_Scalo_1979}
{Miller}, G.~E., \& {Scalo}, J.~M. 1979, \apjs, 41, 513, \dodoi{10.1086/190629}

\bibitem[{{Miller} {et~al.}(2005){Miller}, {Freitag}, {Hamilton}, \& {Lauburg}}]{Miller_2005}
{Miller}, M.~C., {Freitag}, M., {Hamilton}, D.~P., \& {Lauburg}, V.~M. 2005, \apjl, 631, L117, \dodoi{10.1086/497335}

\bibitem[{{Naoz} {et~al.}(2022){Naoz}, {Rose}, {Michaely}, {Melchor}, {Ramirez-Ruiz}, {Mockler}, \& {Schnittman}}]{Naoz_2022}
{Naoz}, S., {Rose}, S.~C., {Michaely}, E., {et~al.} 2022, \apjl, 927, L18, \dodoi{10.3847/2041-8213/ac574b}

\bibitem[{{Pan} {et~al.}(2021){Pan}, {Lyu}, \& {Yang}}]{Pan2021}
{Pan}, Z., {Lyu}, Z., \& {Yang}, H. 2021, \prd, 104, 063007, \dodoi{10.1103/PhysRevD.104.063007}

\bibitem[{{Pan} \& {Yang}(2021)}]{Pan_Yang21}
{Pan}, Z., \& {Yang}, H. 2021, \prd, 103, 103018, \dodoi{10.1103/PhysRevD.103.103018}

\bibitem[{{Perets} {et~al.}(2007){Perets}, {Hopman}, \& {Alexander}}]{Perets2007}
{Perets}, H.~B., {Hopman}, C., \& {Alexander}, T. 2007, \apj, 656, 709, \dodoi{10.1086/510377}

\bibitem[{{Peters}(1964)}]{Peters_1964}
{Peters}, P.~C. 1964, Physical Review, 136, 1224, \dodoi{10.1103/PhysRev.136.B1224}

\bibitem[{{Press} {et~al.}(1992){Press}, {Teukolsky}, {Vetterling}, \& {Flannery}}]{num_recipes_1992}
{Press}, W.~H., {Teukolsky}, S.~A., {Vetterling}, W.~T., \& {Flannery}, B.~P. 1992, {Numerical recipes in C. The art of scientific computing}

\bibitem[{{Preto} \& {Amaro-Seoane}(2010)}]{Preto_Amaro_Seoane_2010}
{Preto}, M., \& {Amaro-Seoane}, P. 2010, \apjl, 708, L42, \dodoi{10.1088/2041-8205/708/1/L42}

\bibitem[{{Qunbar} \& {Stone}(2023)}]{Qunbar_Stone_2023}
{Qunbar}, I., \& {Stone}, N.~C. 2023, arXiv e-prints, arXiv:2304.13062, \dodoi{10.48550/arXiv.2304.13062}

\bibitem[{{Rauch} \& {Tremaine}(1996)}]{Rauch_Tremaine_1996}
{Rauch}, K.~P., \& {Tremaine}, S. 1996, \na, 1, 149, \dodoi{10.1016/S1384-1076(96)00012-7}

\bibitem[{{Raveh} \& {Perets}(2021)}]{Raveh_2021}
{Raveh}, Y., \& {Perets}, H.~B. 2021, \mnras, 501, 5012, \dodoi{10.1093/mnras/staa4001}

\bibitem[{{Rom} {et~al.}(2024){Rom}, {Linial}, {Kaur}, \& {Sari}}]{Rom_24}
{Rom}, B., {Linial}, I., {Kaur}, K., \& {Sari}, R. 2024, arXiv e-prints, arXiv:2406.19443, \dodoi{10.48550/arXiv.2406.19443}

\bibitem[{{Salpeter}(1955)}]{Salpeter_1955}
{Salpeter}, E.~E. 1955, \apj, 121, 161, \dodoi{10.1086/145971}

\bibitem[{{Sari} \& {Fragione}(2019)}]{Sari_Fragione_2019}
{Sari}, R., \& {Fragione}, G. 2019, \apj, 885, 24, \dodoi{10.3847/1538-4357/ab43df}

\bibitem[{{Sch{\"o}del} {et~al.}(2020){Sch{\"o}del}, {Nogueras-Lara}, {Gallego-Cano}, {Shahzamanian}, {Gallego-Calvente}, \& {Gardini}}]{Schodel_2020}
{Sch{\"o}del}, R., {Nogueras-Lara}, F., {Gallego-Cano}, E., {et~al.} 2020, \aap, 641, A102, \dodoi{10.1051/0004-6361/201936688}

\bibitem[{{Sigurdsson} \& {Rees}(1997)}]{Sigurdsson_1997}
{Sigurdsson}, S., \& {Rees}, M.~J. 1997, \mnras, 284, 318, \dodoi{10.1093/mnras/284.2.318}

\bibitem[{{Teboul} {et~al.}(2024){Teboul}, {Stone}, \& {Ostriker}}]{Teboul_2024}
{Teboul}, O., {Stone}, N.~C., \& {Ostriker}, J.~P. 2024, \mnras, 527, 3094, \dodoi{10.1093/mnras/stad3301}

\bibitem[{Vasiliev(2017)}]{Vasiliev_17}
Vasiliev, E. 2017, The Astrophysical Journal, 848, 10, \dodoi{10.3847/1538-4357/aa8cc8}

\bibitem[{{Weissbein} \& {Sari}(2017)}]{Weissbein_Sari_2017}
{Weissbein}, A., \& {Sari}, R. 2017, \mnras, 468, 1760, \dodoi{10.1093/mnras/stx485}

\end{thebibliography}
\bibliographystyle{aasjournal}
  
 \end{document}